\documentclass[12pt]{article}
\usepackage{geometry}
\usepackage{amsfonts}
\usepackage{amsmath}
\usepackage{amssymb}

\usepackage{mathrsfs}
\usepackage{wasysym}

\usepackage{bbm}
\usepackage{bm}
\usepackage{graphicx}
\usepackage{hyperref}
\usepackage[usenames,dvipsnames]{color}

\newcommand{\st}[1]{\text{\tiny \rm #1}}

\def\e{\emph}
\def\bq{\begin{equation}}
\def\ee{\end{equation}}

\def\nb{$N$-body problem }
\def\nbn{$N$-body problem}

\def\tb{3-body problem~}
\def\tbn{3-body problem}

\def\id{, i.e., }

\def\Rcm{{\bf R}^\st{\,cm}}
\def\shs{\mathsf{S}}

\def\Sim{\mathsf{Sim}}

\renewcommand{\d}{{\rm d}}

\newcommand{\jan}{{\sf{J}}}
\newcommand{\janp}{{\sf{j}}}

\hypersetup{
colorlinks=true,         
linkcolor=blue,          
citecolor=red,        
urlcolor=Violet            
}

\title{\vspace{-1.5cm}\bf Entropy and the Typicality of Universes}

\author{{Julian Barbour,$^{1,2}$
Tim~Koslowski,$^{3}$ and Flavio~Mercati,$^4$
{\rm }
}\vspace{12pt} \\
\it \small $^1$College Farm, South Newington, Banbury, Oxon, OX15 4JG UK,\\
\it \small $^2$Visiting Professor in Physics at the University of Oxford, UK.\\
\it \small  $^3$ University of New Brunswick, Fredericton, NB, E3B 5A3 Canada,\\
\it \small $^4$Perimeter Institute for Theoretical Physics, 31 Caroline Street North,\\
\it \small Waterloo, ON, N2L 2Y5 Canada,}
\date{}

\begin{document}

\maketitle

\begin{abstract}
The universal validity of the second law of thermodynamics is widely
attributed to a finely tuned initial condition of the
universe. This creates a problem: why is the universe atypical?
We suggest that the problem is an artefact created by
inappropriate transfer of the traditional concept of entropy to
the whole universe. Use of what we call the relational
$N$-body problem as a model indicates the need to
employ \emph{two} distinct entropy-type concepts to describe the
universe. One, which we call \emph{entaxy}, is novel. It is
scale-invariant and \emph{decreases} as the observable universe
evolves. The other is the algebraic sum of the dimensionful
entropies of branch systems (isolated subsystems of the
universe). This conventional additive entropy increases. In our model, 
the decrease of entaxy is fundamental and makes possible the emergence of branch systems and their increasing entropy. We have previously shown that all
solutions of our model divide into two halves at a unique `Janus point'
of maximum disorder. This constitutes a common past for two futures each with its own gravitational arrow of time. We now show that these arrows are expressed through the formation of branch systems within which conventional entropy increases. On either side of the Janus point, this increase is in
the same direction in every branch system. We also show that it is only
possible to specify unbiased solution-determining data at the
Janus point. Special properties of these `mid-point data' make it
possible to develop a rational theory of the typicality of
universes whose governing law, as in our model, dictates
the presence of a Janus point in every solution. If our
self-gravitating universe is governed by such a law, then the
second law of thermodynamics is a necessary direct consequence of
it and does not need any special initial condition.
\end{abstract}

\maketitle

\newpage


\section{Introduction}

\label{SecIntro}

Entropic arguments are very effective when applied to confined systems\,\footnote{\,We  distinguish \e{dynamically closed} systems, on which no external forces act, from \e{confined} systems such as particles in a box. The former can expand freely, as is the case with the universe.} in equilibrium. However, the universe is unconfined and far from equilibrium. Is it then correct to apply conventional entropy concepts to it? We think not. New issues come into play when the universe, which by definition is a unique system, is considered.

We first argue that the universe's scale and shape degrees of freedom must be treated differently. 
Scale degrees of freedom and time intervals are useful to characterize subsystems, for which the rest of the universe provides a reference frame with respect to which length and time units can be defined. They lose meaning, however, when applied to the whole universe, which by definition admits nothing external to it. Any property of the whole universe, including functions used to quantify its typicality, must be expressed in terms of scale-invariant ratios of quantities within the universe together with the changes of these ratios. If this is done, the principles of spatial and temporal relationalism \cite{FlaviosSDtutorial} are realized.


Next, the specification of \e{initial} conditions is suspect: in many unconfined systems governed by time-symmetric laws \e{all} solutions divide into two qualitatively similar halves at a unique `central' point. We call it the \e{Janus point} $\janp$ (after the Roman god that looks in two opposite directions at once) and say that the corresponding solutions are \e{Janus-point solutions}. To our knowledge, the existence and potential significance of such solutions have never been noted in the voluminous literature on the second law of thermodynamics and its application to the universe. 

As an example of the difference that they can make, we showed in \cite{BKM3} that in self-gravitating Newtonian systems subject to natural restrictions \e{all} solutions have a Janus point\,\footnote{\,In \cite{BKM3}, this coinage had not yet occurred to us.} and contain arrows of time, expressed through the growth of a scale-invariant quantity which we call \e{complexity},\footnote{\,It measures the clustering of the particles in the system} that arise of dynamical necessity. Contrary to a widespread belief, this result shows that all the solutions of time-reversal symmetric dynamical systems can exhibit \e{time-asymmetric} behaviour for internal observers, who must be on one or the other side of the Janus point. The arrows of time are therefore not due to any special initial condition (or `past hypothesis' \cite{albert2009time}) but are rather a direct consequence of the law of the universe. This result suggests that previous discussions of the second law of thermodynamics have failed to note the one fact -- dynamically enforced Janus-point solutions -- that could be the true explanation for the universal validity of the second law in our universe. We here present more support for this view. 

First, we show that when a Janus point does exist it is the only point at which solution-determining data can be specified without bias. We call such data `mid-point data'. Moreover, the model that we used in \cite{BKM3} and call the \e{relational} $N$-\e{body problem} has properties which make it a good `toy' universe and possesses mid-point data with remarkable properties. They are \e{scale-invariant and non-redundant.} In addition, they stand in a one-to-one relation to the points of a geometrical structure known as a \e{projectivized cotangent bundle}, which we denote by $PT^*{\sf S}$ (the $\sf{S}$ stands for \e{shape space}, a key notion in our theory). We detail below the properties of  $PT^*{\sf S}$ and show that it carries a \e{natural bounded measure}. 

This fact plays a central role in our paper because it enables us to develop a mathematically rigorous theory of \e{the typicality of universes}. The point is that, at least in the relational \nbn, we know the law that governs the universe but not the probability with which any particular solution will be realized. Under these conditions, as argued in \cite{GHS}, the only rational choice one can make is to assume all solutions have \e{equal probability} of being realized. This is Laplace's well-known \e{principle of indifference}. If this is to be applied consistently in a situation in which there is a continuum of possible solutions, the solution space must be endowed with a bounded measure. This is exactly what we possess by virtue of the natural measure on $PT^*{\sf S}$, whose points map one-to-one onto the solution space of the theory. As we will show, the existence of the Janus point, the scale invariance of the mid-point data and the bounded measure associated with them gives our theory three distinct advantages compared with the situation in \cite{GHS}, where the corresponding measure is infinite and there is no Janus point. The three advantages lead us to an entirely novel overall picture of the universe.

Indeed, the full development of our theory shows that, at least in our model, the universe must be described by \e{two} distinct entropy-type quantities. Both are based on Boltzmann's great insight: entropy is a count of microstates compatible with a given value of a state function defined on the phase space of the theory. For the universe as a whole, we introduce the notion of \e{entaxy} (from the Greek for `towards order'). It is a scale-invariant count of microstates compatible with given values of the scale-invariant complexity. In any solution, it is maximal in the neighbourhood of the Janus point and \e{decreases} with increasing distance in either direction from the Janus point. The very fact that the entaxy is decreasing reflects the breakup of an initially (at the Janus point) very uniform universe into gravitationally bound clusters (branch systems) for which an increasing conventional (dimensionful) Boltzmann entropy is defined as a (microcanonical) count of microstates compatible with the (dimensionful) values of the energy and angular momentum of the clusters. Thus, there is an \e{increasing} entropy of the universe. It is obtained as the algebraic sum of the bound-cluster entropies, the overwhelming majority of which increase in either direction away from the Janus point. In the framework of our theory, the entaxy is clearly the more fundamental quantity: it must decrease, and thereby creates the conditions under which branch systems can form and increase their entropy.

As a proof of principle, we consider as an example the \tb and show that our theory is \e{predictive}: solutions that have more probable mid-point data have asymptotic behaviour which is characteristically different from that of the solutions with less probable mid-point data. Thus, observers within such a universe could in principle establish whether they belong to a typical or an atypical universe. This is a central issue in modern cosmology. Our approach suggests it may be amenable to resolution.

We are also able to show that in all systems with Janus points $\janp$ the scale-invariant asymptotic behaviour that develops with increasing distance in either direction from $\janp$ is subject to dynamical attractors of the shape degrees of freedom. This leads to strong correlations reminiscent of the behaviour associated with retarded potentials in wave processes. This  suggests that they too are a dynamical necessity and not the result of a special condition `in the past'.

In this paper, we have advisedly restricted attention to the relational \nbn. Its simplicity enables us to deduce all our conclusions with rigour and present the broad principles of our theory
in a transparent manner. Of course, the real question is whether the theory is appropriate for \e{the universe}. As we note in our conclusions, this will very likely be so if the big-bang is not the beginning of time but instead a Janus-point that hides from our view a qualitatively similar twin universe on its other side in which the internally observed direction of time is the opposite to ours.

\section{Shape, Size and Entropy\label{SSE}}

 \begin{figure}[b!]
\center\includegraphics[width=\textwidth]{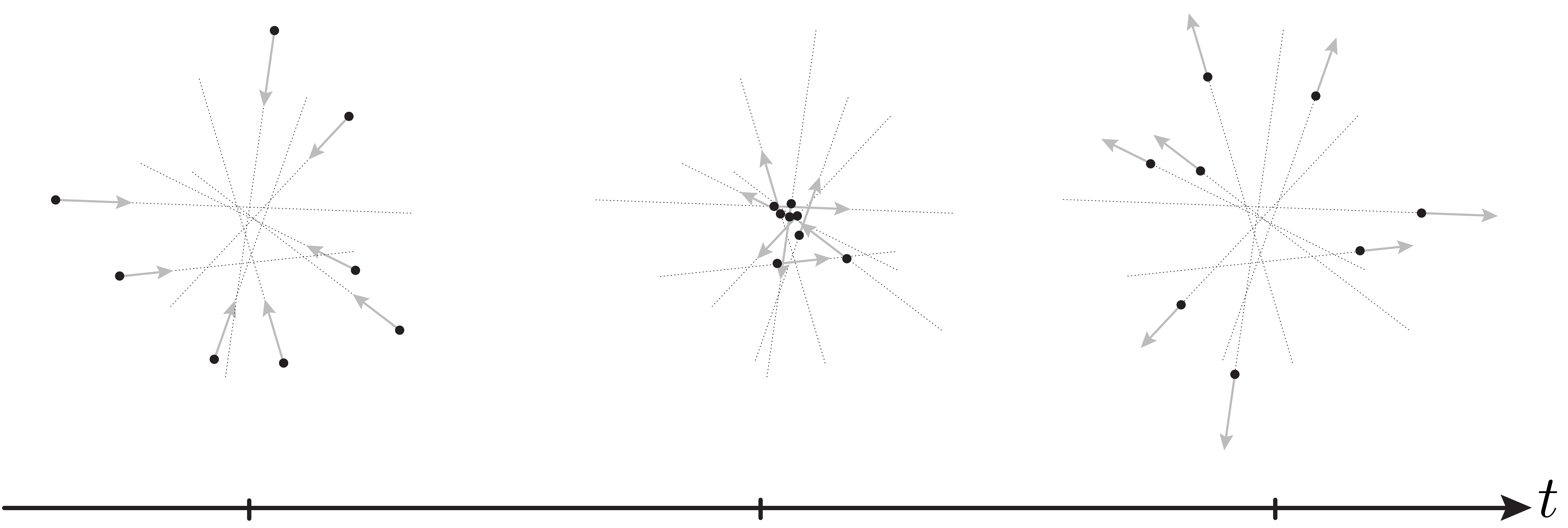}
\caption{\small A planar solution of an  8-body system of inertial particles. Each particle moves at a constant speed in a straight line (from the perspective of an inertial reference frame and with Newtonian time~\cite{FlaviosSDtutorial}), ${\bf r}_a = {\bf r}_a(t_0) + {\bf v}_a (t-t_0)$. At both time boundaries the system tends to two parity-conjugate shapes, which are completely determined by the velocity vectors $ {\bf v}_a$ and do not depend on the initial (or better mid-point, see below) coordinates. Because the system is in two dimensions and not three, the parity conjugation actually leaves the shapes identical.
\label{InertialFigure}}
\end{figure}

Point particles moving inertially model the simplest possible universe. If the particles were small elastic spheres, they would model an ideal gas. If confined \e{and in equilibrium}, its entropy  would be well defined as would changes of it under controlled adiabatic manipulation. But if unconfined, what can be said about the system in itself? If ${\bf r}_a^\st{\,cm}$ is particle $a$'s position relative to the centre of mass $\Rcm$, the moment of inertia is
\bq
I_\st{cm}=\sum_a^N m_a\,{\bf r}_a^\st{\,cm}\cdot{\bf r}_a^\st{\,cm}\equiv {1\over M}\sum_{a<b}m_a\,m_b\,r_{ab}^{\,2},~~M=\sum_a^Nm_a,~~r_{ab}=\|{\bf r}_b-{\bf r}_a\|, \label{cm}
\ee
and the root-mean-square length
\bq
\ell_\st{rms}={1\over M^{\,1/2}}\sqrt{I_\st{cm}}\equiv{1\over M}\sqrt{\sum_{a<b}m_a\,m_b\,r_{ab}^{\,2}}\label{rms}
\ee
is the degree of freedom (dof) that measures the system's size. An external frame defines position and orientation dofs. Truly intrinsic to the system are its shape dofs.

Consider now a system of inertial particles: $N$ uninteracting particles, corresponding to an ideal gas in empty space. Whatever initial positions and momenta we attribute to the $N$ particles, they will fly off undisturbed each along its own trajectory, dispersing to infinity. The moment of inertia and $\ell_\st{rms}$ will grow unboundedly large. The asymptotic configuration will have $\ell_\st{rms} = \infty$ and a shape determined only by the momenta. The information about the initial positions will be completely erased, and the cm coordinates ${\bf r}_a^\st{\,cm}$ and momenta ${\bf p}^a_\st{\,cm}$ will become more and more correlated as  $\ell_\st{rms}\rightarrow\infty$: each particle asymptotes to recession from $\Rcm$; the angles $\measuredangle\,\{ {\bf r}_a^\st{\,cm},{\bf p}^a_\st{\,cm}\}\rightarrow 0$ and any two particles will tend to separate with speed $v_{ab} \propto r_{ab}$ as in Hubble-type cosmological expansion.

We now note that our system of inertial particles has time-reversal symmetric equations of motion: if we evolve those same initial conditions backwards in time (or, equivalently, reverse all the initial velocities), we obtain the same asymptotics at $t\to -\infty$, with the difference that the asymptotic shape is the parity-reversed image of the other. Moreover, $\ell_\st{rms}$ tends to $\infty$ both at $t\to \infty$ and $t\to -\infty$. It is easy to see that $\ell_{rms}^{\,2}$ is a positive quadratic form in $t$, and therefore reaches a unique minimum in each solution. For this reason, systems of inertial particles are the simplest with a Janus point $\janp$: $\ell_\st{rms}$ always passes through a unique minimum, and $\ell_\st{rms}\rightarrow\infty$ as $t\rightarrow\pm\infty$. It would clearly be inappropriate to try to specify `initial' conditions freely at either $t\rightarrow\pm\infty$ limit due to the tight correlations between ${\bf r}_a^\st{\,cm}$ and ${\bf p}^a_\st{\,cm}$.

Entropy growth is often described qualitatively as passage from an atypical (ordered) to a typical (disorded) state. Let us consider this  in the light of the observations we have just made. If we specify uncorrelated initial data for a system of inertial particles, the state corresponding to those data must be near the Janus point of the solution they generate -- at any distance from $\janp$ the particle motions become correlated, as we have just noted. Thus, the state of the system of inertial particles becomes less typical going away from $\janp$, and its `arrows of typicality growth' point \emph{towards} $\janp$. This is typical of all unconfined systems with Janus-point solutions. 

It is illuminating to compare this behaviour with that of an ideal gas in a confined space. In contrast to the system of unconfined inertial particles, the accessible phase space of the system then has a \e{bounded measure} and is subject to Poincar\'e recurrences: its phase-space point will return arbitrarily often arbitrarily close to any point previously visited. Although the time between close returns is extremely long (for a mole of gas much longer than the age of the universe), this behaviour is inevitable due to the fact that the accessible phase space is bounded. Because of the recurrences, the (Boltzmann) entropy $S_\st{B}$ of the system must fluctuate. For most of the time, $S_\st{B}$ will remain very close to its maximum, which corresponds to thermal equilibrium (heat death). However, there will inevitably be occasional fluctuations to lower values of $S_\st{B}$. These will be ever rarer, the deeper the dips in $S_\st{B}$. 

The plot of $S_\st{B}$ against the time $t$ will exhibit qualitative time-reversal symmetry, reflecting the time-reversal symmetry of the law that governs the system. In a famous letter to \e{Nature}~\cite{boltzmann1895certain}, Boltzmann conjectured that we find ourselves in a world in which entropy increases because we happen to be on one side of a deep Poincar\'e dip and define the direction of time as the direction away from the dip, towards equilibrium and heat death. Intelligent beings on the other side of the dip will find that their time flows in the opposite direction, again to equilibrium and heat death. Since typicality increases with entropy, in such a system the arrows of typicality growth point \e{away} from the dip. Because the behaviour is symmetric about the dip, there is qualitative time-reversal symmetry about it just as there is for the unconfined system of inertial particles. The crucial difference is, however, that the arrows of typicality growth in the confined system point \emph{away} from the dip, and not toward the dip as the Janus-point system. Moreover, whereas there is a single Janus point in the unconfined system, there are infinitely many dips in the confined system.

The origin of the `towards--away' mismatch and also the fact that in one case there are infinitely-many dips but in the other just one special point is obvious. If a system of elastic inertial particles with $\ell_\st{rms}$ increasing freely is abruptly confined in an elastic box, collisions of the particles with the box and each other will rapidly reverse its atypicality growth, and a statistically typical equilibrium state with Poincar\'e recurrences will be established. Prevention of $\ell_\st{rms}$ growth forces the shape dofs to equilibrate and explains both of the mismatches.

Natural processes whose time reverse is never observed are said to be irreversible and associated with entropy growth. An example sometimes given (\cite{Davies}, p. 33) is that ``a gas will explode into a vacuum, but will never spontaneously implode into a smaller volume''. This is certainly true \e{within} the universe, but a universe made only of $N$ inertial particles `implodes' from $t=-\infty$ to $\janp$ and then `explodes' to $t=+\infty$ for either nominal time direction.\,\footnote{\,\label{retarded}A half-solution, from $\ell_\st{rms}=\infty$ to $\janp$, has a seemingly implausible initial condition at $\ell_\st{rms}=\infty$ like an advanced-wave solution in electrodynamics. This suggests that the invariable observation of retarded waves in the universe is due to its having a Janus point $\janp$ and our observations being made in an asymptotic region.}

Now consider the role size plays in ideal-gas entropy. Up to an additive constant
\bq
S=Nk\,\textrm{log}\,(VT^{\,3/2}),\label{ent}
\ee
where $k$ is Boltzmann's constant, $V$ the volume, and $T$ the temperature. Under reversible adiabatic expansion,\footnote{\,Adiabatic expansion of a gas in a piston is reversible because the overall entropy of the system+environment does not change, even though the gas increases its entropy in the process.} $S$ increases because the system constantly re-equilibrates, thereby maximizing the entropy subject to new constraints. Note that, the particles and container being elastic, $T$ remains constant; $S$ increases because $V$ does. But without the constraint-enforced shape equilibration, calculating an entropy is problematic. The Gibbs $H$ function is constant under Hamiltonian evolution: it is extremalization of $H$ subject to constraints that leads to experimentally verifiable entropy values~\cite{Jaynes-BoltzmannVsGibbs}. The Boltzmann entropy $S_\st{B}$, based on a one-particle distribution in a six-dimensional phase space and obtained from the Gibbs $H$ by integrating over the degrees of freedom of all but one of the particles, can increase. However, it inherits the ensemble nature of the Gibbs $H$ function, and it is ensemble spreading that leads to the appearance of typicality growth.\,\footnote{\,Note also that the Boltzmann entropy does not in general (unless one is considering an ideal gas) lead to the correct empirically determined equilibrium thermodynamic relations~\cite{Jaynes-BoltzmannVsGibbs} because its one-particle distribution function washes out correlations taken into account in the Gibbs $H$ function based on the $N$-particle distribution function.} In reality, every single microsolution within the ensemble tends with increasing distance from $\janp$ to an ever more atypical, increasingly correlated state. This is quite unlike a confined gas as it approaches equilibrium, in which virtually all correlations are being destroyed in every microsolution.

\begin{figure}[h!]
\center\includegraphics[width=0.5\textwidth]{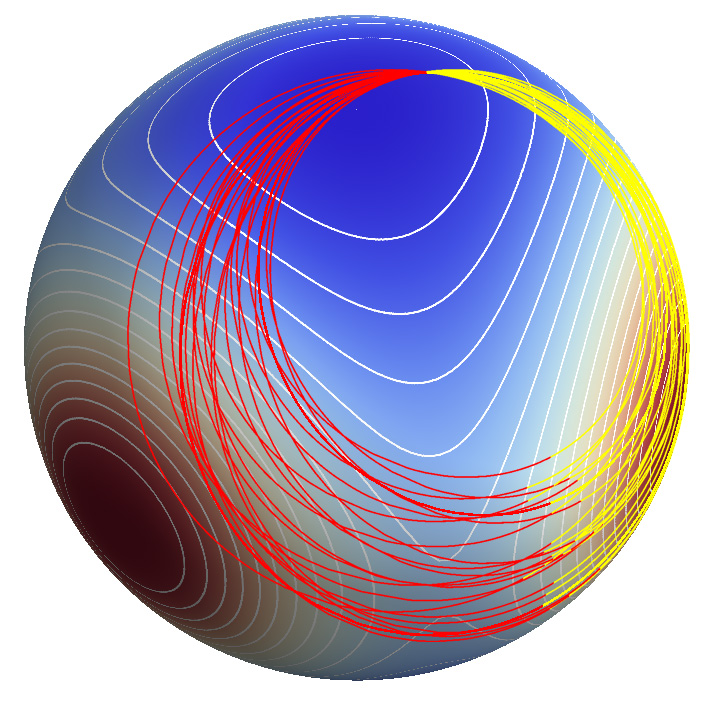}
\caption{\small 
This diagram shows a bundle of 30 solutions of the inertial \emph{planar} 3-body problem in the space of scale-invariant configurational degrees of freedom, the \emph{shape sphere} (defined in the next section and explicitly constructed in Appendix A). Planar means that the angular momentum is perpendicular to the plane of the 3 bodies, and therefore the motion unfolds on a fixed plane. Each solution is divided in two (a red and a yellow part), at each side of its Janus point. All solutions have  the same momenta but the mid-point coordinates are normally distributed around a central shape, with a variance of $\sim$25\%.  In this way one gets a distribution in shape space whose evolution, at both asymptotic ends, converge to a single shape (as noted in the caption to Fig.~\ref{InertialFigure}, in two dimensions parity-conjugate triangles are identical), completely determined by the momenta. The presence of dynamical attractors is here manifest; later, when we add gravitational interaction to inertial motion, we will find an even more impressive manifestation of such attractors.
\normalsize} \label{DynamicalAttractors}
\end{figure}

Moreover, phase-space volume is a product of size and shape factors. In unconfined Janus-type systems, the sizes in any Gibbs ensemble can spread freely, so the size factor will grow in both time directions. To keep $H$ constant, the shape factor must decrease. The shape dofs are therefore subject to \e{dynamical attractors}, as demonstrated clearly in Fig.~\ref{DynamicalAttractors}. Their existence is hidden unless scale is factored out.

Finally, overall size, unlike dimensionless shape dofs, only has meaning relative to an external scale, which does not exist for a universe. The justification for saying that our universe is expanding is based on ratios, in the first place of wavelengths in red shifts. Observations of them make it meaningful to say that the universe at a given epoch is twice as large as it was at some earlier epoch, but not to ascribe size to any specific epoch. Relative scale is meaningful but absolute scale is not.

In the light of these considerations, of which we have found little recognition in the 
literature,\,\footnote{\,Sloan~\cite{Sloan:2014qy} points out that in cosmology the presence of the scale factor in the case of a minimally coupled scalar field hides the presence a shape attractor.}  we conclude that entropic arguments require revision when applied to the universe. We can summarize our position as follows: in any dynamically closed but unconfined system that can increase arbitrarily in size (in the sense defined above) and whose solutions have a Janus point, judgements as to the growth of typicality need to be based on the behaviour of the shape dofs, which are the only ones that are directly observable. The `invisible' (and single) scale dof does play an important role in `allowing' the shape dofs to change their overall typicality but should not be included in any calculation of typicality.

Thus, the critical issue in the definition of entropy for the universe is, as we see it, this: as a physical system, is the universe confined or unconfined?\,\footnote{\,Let us stress that our notion of \emph{confined} is not equivalent to \emph{finite}, or \emph{compact}. These are two completely unrelated concepts.} If the latter, which seems intuitvely natural, how can we define for it a bounded measure, without which an entropy-type count of microstates for a given macrostate is ill defined?

\section{The Relational $N$-Body Universe}

We obtain a model of the universe that is much more realistic than inertial particles by introducing Newtonian gravitational interaction between particles of mass $m_a, a=1 \dots N$. 
It is a model ideally suited for our purposes: simple but nontrivial enough to illustrate our main points. It is also valuable in that it brings out the difference between gravitational and nongravitational forces.

What we mean by a \e{relational} model is one that makes no use of extrinsic elements. The dimensionless mass ratios $m_a/M$ and the dimensionless inter-particle separations $s_{ab}:= r_{ab}/\ell_\st{rms}$ are intrinsic quantities. If $N\ge 5$ these are not all independent; the instantaneous \e{shape} is determined by $3N-7$ independent combinations of the $s_{ab}$. The extrinsic elements in Newtonian dynamics enter through an inertial frame of reference. This provides an external time parameter $t$, a centre-of-mass velocity,\,\footnote{\,By Galilean relativity, this cannot be observed in the behaviour of the $s_{ab}$ and is of no concern.}  an orientation of axes and a scale.  None of the extrinsic elements are directly observable. 
The $3N$ particle coordinates ${\bf r}_a$ in an inertial frame are therefore heterogeneous, mixing instrinsic and extrinsic degrees of freedom:
\begin{itemize}
\item three locate the position of the system in the frame, e.g., the centre of mass 
\begin{equation}
\Rcm = \sum_{a=1}^N m_a {\bf r}_a / M, \qquad M=\sum_{a=1}^N m_a
\end{equation}
\item three fix its orientation, e.g., the orientation\,\footnote{\,Notice that the orientation of a vector involves just two degrees of freedom, but the orientation of a system of three orthogonal vectors involves three dofs.} of the 3 eigenvectors of the centre-of-mass moment of inertia tensor (fixing the `principal axes of inertia')\,\footnote{\,This is just an illustrative example. It is well known that there are spherically- or cylindrically-symmetric configurations whose principal axes of inertia are degenerate and cannot be used to fix an orientation in space. However, these configurations are special (measure zero).}
\begin{equation}
\mathbbm{I}_\st{cm} =\sum_{a=1}^N m_a\left( \mathbbm{1} \, {\bf r}_a^\st{cm}\cdot{\bf r}_a^\st{cm} - {\bf r}_a^\st{cm}\otimes{\bf r}_a^\st{cm} \right), \qquad {\bf r}_a^\st{cm} = {\bf r}_a - {\bf r}_\st{cm},
\end{equation}
\item  one fixes its size, measured for example by the moment of inertia (\ref{cm}):
\bq
I_\st{cm}= \frac 1 2 {\rm tr} ( \mathbbm{I}_\st{cm} ) =  \sum_{a=1}^N m_a\,{\bf r}_a^\st{cm}\cdot{\bf r}_a^\st{cm}\equiv{1\over M}\sum_{a<b}m_am_b\,r_{ab}^2,~~r_{ab}= \|{\bf r}_b-{\bf r}_a \|,\label{CM}
\ee
\item the remaining $3N-7$ are the \e{relational, scale invariant} data~\cite{barbour:scale_inv_particles,barbour:nature,FlaviosSDtutorial} (in short, \e{shape data}). They are angles and ratios of the separations $r_{ab}$ between the particles. Only these are observables.
\end{itemize}
None of the extrinsic elements are directly observable. To establish how they affect what is observable, we define \e{shape space} ${\sf S} = {\sf Q} / {\sf Sim}$ by quotienting the Newtonian configuration space ${\sf Q}  \sim \mathbbm{R}^{3N}$ by the similarity group $\Sim$:
\bq
 {\sf Sim} := \left\{  {\bf r}_a \sim {\bf r}'_a ~ \text{iff} ~  \exists ~ {\bm \theta} \in {\mathbbm R}^3, \, \Omega \in SO(3),  \varphi >0 ~ / ~ {\bf r}'_a = \varphi ( \Omega \, {\bf r}_a + {\bm \theta}) \right\}.
\ee
Any Newtonian solution can be projected to a curve in $\shs$. This is simply a succession of shapes without the earlier--later ordering provided by $t$. In the \tbn, the simplest non-trivial \nbn, two angles, $\alpha$ and $\beta$, determine the instantaneous shape of the triangle that the particles form. The direction of change is $\textrm d\alpha/\textrm d\beta$. In a purely intrinsic dynamics, $\alpha,\beta,\textrm d\alpha/\textrm d\beta$ should determine the evolution $\alpha  = \alpha(\beta)$. 
This is not so for a generic Newtonian solution because five dimensionless quantities of extrinsic origin play a role. They are: 
\begin{itemize}
\item the \emph{direction} of the cm angular momentum vector ${\bf L}=\sum_a{\bf r}_a^\st{cm} \times{\bf p}^a_\st{cm}$ (two dofs), 

\item the ratio $T_\st{cm}/V_\st{New}$ of the cm kinetic energy to the Newtonian potential energy,  

\item the ratio $T_\st{R}/T_\st{S}$ between the rotational and shape contributions to 
$T_\st{cm}$,\,\footnote{\,The \e{velocity decomposition theorem}~\cite{SaariBook} states that at any instant $T_\st{cm}=T_\st{R}+T_\st{D}+T_\st{S}$.}

\item the ratio $T_\st{D}/T_\st{S}$ between the dilatational and shape contributions to $T_\st{cm}$.
\end{itemize}
The values of the five dimensionless quantities are not encoded in the shape $s$ and direction $d$ of the curve in $\shs$ but in the way the curve \e{bends} at $s$. Intrinsic $s,d$ initial data fail to determine the intrinsic evolution (they fall short by five for all $N\ge 3$).

The three data related to the angular momentum ${\bf L}$ (two angles and the ratio $T_\st{R}/T_\st{S}$) are present because orientation is absolute in Newtonian theory. They cease to play a role if ${\bf L}=0$. Similarly, if the total energy $E$ vanishes, $E=0$, the absolute time plays no role and the ratio $T_\st{cm}/V_\st{New}$ must have the value $-1$. Otherwise, it will change along the curve in $\shs$ and can have any value at an initial shape $s$. It is important that the conditions $E={\bf L}=0$ are conserved by Newtonian dynamics. This means we can consistently eliminate the effect of four of the five extrinsic quantities.

We now come to the most critical point: whereas the angular momentum ${\bf L}$ is conserved in the \nbn, the \e{dilational momentum}\,\footnote{\,This name for $D$, which is the generator of dilatations, was introduced in~\cite{barbour:scale_inv_particles}. It has the same dimensions as angular momentum, the generator of rotations. There is no name for $D$ in the $N$-body literature; it is usually denoted by $J$, probably for Jacobi.} $D=\sum_a{\bf r}_a^\st{cm}\cdot{\bf p}^a_\st{cm}$ is not. (Unlike ${\bf L}$, $D$ does not commute with the Hamiltonian.) 
Because of this, the ratio $T_\st{D}/T_\st{S}$ of the dilatational and shape kinetic energies cannot vanish throughout the evolution; even when $E={\bf L}=0$, a shape and direction in $\shs$ are in general insufficient to determine the evolution curve in $\shs$. What one finds is that for any point $s$ in $\shs$ and any direction $d$ at $s$ there is a one-parameter family of projected Newtonian solutions through $s$ whose tangents all have the same direction $d$. With increasing distance from $s$, they `peel off' from the common direction $d$ into manifestly distinct solutions. This characteristic behaviour is due to the role that relative (as opposed to absolute) scale plays in Newtonian gravity, as we shall now show.

For any potential $V$ homogeneous of degree $k$,\,\footnote{\,A function $f(x_i)$ is homogeneous of degree $k$ if $f(ax_i)=a^kf(x_i)$ for any constant $a>0$.} it is easy to show that 
\bq
\ddot I_\st{cm}=2E-(k+2)V.\label{lj}
\ee
The Newton potential is homogeneous of degree $-1$, so that for it $\ddot I_\st{cm}=2E-V_\st{New}$. Moreover, $V_\st{New}$ is negative-definite, so that if $E\ge 0$
\bq
\ddot I_\st{cm}>0.
\ee

This means that $\ddot I_\st{cm}$ is concave upwards as a function of $t$ and (half) its first derivative,
\begin{equation}
\frac 1 2 \dot I_\st{cm} =\sum_{a=1}^N {\bf r}_a^\st{cm}\cdot{\bf p}^a_\st{cm}=D,
\end{equation}
is a monotonically increasing function (under time reversal $D$ changes sign and therefore is increasing in both nominal time directions).\,\footnote{\,It is an immediate consequence of the existence of this monotonically inceasing (Lyapunov) function that the solutions of the relational \nb cannot be periodic. The system is not ergodic and cannot be subject to Poincar\' e recurrences.} This is a consequence of the \emph{Lagrange--Jacobi relation}. At one unique point $\janp$ we have $D=0$, \e{the Janus point of the system}. The value of $I_\st{cm}$ at $D=0$ is purely nominal since it depends on the unit chosen to measure distance. However, the ratio of the $I_\st{cm}$ values at distinct points in the evolution is physical and can in principle be determined from the curve in $\shs$. This is also true for the instantaneous value of the ratio $T_\st{D}/T_\st{S}$ and the point where both it and $D$ vanish. Overall (absolute) scale has no objective meaning but relative scale does.\,\footnote{\,To create a dynamics of \e{pure shape}~\cite{barbour:scale_inv_particles,FlaviosSDtutorial}, in which one can simultaneously have $D=E={\bf L}=0$ and therefore a point and direction do determine the evolution in $\shs$, the Newton potential must be replaced by one that is homogeneous of degree $-2$.}

Thus, the basic structure of Newtonian \e{evolution} is not scale invariant. It is however remarkable that each individual, observationally distinct evolution curve in $\shs$ corresponding to Newtonian solutions with $E={\bf L}=0$ is uniquely and nonredundantly determined by scale-invariant quantities. This is due to the existence of the single point at which $D=0$. As we have noted, this is a Janus point, and at this point in the evolution there is no dilatational kinetic energy: $T_\st{D}=0$. Thus, at $D=0$ a point and direction in $\shs$ do determine a solution, moreover non-redundantly.\,\footnote{\,As emphasized in a valuable comment to one of us (JB) by Halliwell [J Halliwell, private communication, September 2009], a general phase-space point determines a solution and, redundantly, an initial point on it. The existence of the unique point at which $D=0$ enables us to specify solution-determining data at that point nonredundantly.} We have \e{mid-point data}.\,\footnote{\,In~\cite{CarrollChen} Chen and Carroll contemplated time-symmetric solutions of time-symmetric equations with entropy growing to infinity either side of a unique minimum. Janus-point systems confirm this `middle hypothesis'~\cite{CarrollBook} though, as we will see below, our entropy-type quantity has a maximum, not minimum, at $\janp$. As Carroll~\cite{CarrollBook} notes, it is intuitively desirable that systems with time-reversal symmetric equations should have solutions that are qualitatively time symmetric. In fact, this has long been recognized to be so in confined systems in which infinitely many Poincar\'e recurrences occur. It is strange that, in the arrow-of-time literature, the common occurrence of systems with Janus point $\janp$ and just two qualitatively symmetric branches either side of $\janp$ does not seem to have been noticed. Besides inertial systems and the relational \nb (the \nb with $E={\bf L}=0$), two other obvious examples are the \nb with replusive gravity and (liquid or electromagnetic) waves with compact support.} 
When working in phase space, we will call the intersection of the hypersurfaces $E={\bf L}={\bf P} = 0$\,\footnote{\,${\bf P} = \sum_a {\bf p}^a =0$ is the total linear momentum of the system.} and $D=0$ the \e{Janus surface}, which we will denote by the symbol $\jan$.

The manner in which these mid-point data are scale invariant is worth spelling out: for $N$ particles, $3N-7$ numbers fix the shape $s$ and $3N-8$ the direction $d$ (in any space, a direction is fixed by one less number than a point). The conditions $E={\bf L}=0$, which fix the class of solutions considered appropriate for a dynamically closed universe, and the condition $D=0$, which fixes the point at which the mid-point data are specified, are scale invariant. Since we regard as objectively different only those Newtonian solutions that map to distinct curves in shape space $\shs$, we conclude that our space of solutions has $6N-15$ dimensions. Finally, we pointed out earlier that through every shape $s\in {\sf S}$ and for every direction in $\shs$ at $s$ there passes a one-parameter family of solutions all tangent to $d$. Just one of these solutions has $D=0$ at $s$.

The non-redundance of mid-point data, which leads to the odd-dimensionality of the solution space, is due to \e{dynamical similarity}, which arises because the Newton potential is homogeneous \cite{LL1, BKM2}. By rescaling all the initial (or better `mid-point') momenta ${\bf p}^a \to k\,{\bf p}^a$ by a common constant factor $k$, one obtains a solution that is congruent (geometrically similar) to the original one, namely ${\bf r}_a \to k^{-2}\,{\bf r}_a$. If parametrized with Newtonian time (which can be defined relationally as the \emph{ephemeris time} of~\cite{barbourbertotti:mach,FlaviosSDtutorial}), this solution runs at a different pace, which is just a rescaling $t \to k^{-3}\,t$.
So the two solutions differ by an unobservable global rescaling and an equally unobservable time rescaling. They are indistinguishable to observers living in such solutions, who have access only to ratios between space and time intervals within their universes.

Although $\shs$ aids intuition, calculations are more convenient in the Newtonian phase space $\Gamma$.
This contains the following seven constraint hypersurfaces:
\bq
{\bf P} = \sum_{a=1}^N {\bf p}^a =0, \qquad {\bf L}  = \sum_{a=1}^N \,{\bf r}_a\times{\bf p}^a = 0, \qquad D  = \sum_{a=1}^N \,{\bf r}_a\cdot {\bf p}^a =0, 
\ee
According to the classification of constraints introduced by Dirac~\cite{dirac:lectures,HenneauxTeitelboim}, these seven constraints form a 
 first-class system; the intersection ${\sf J} \subset \Gamma$ of all the mentioned hypersurfaces is a $3N-7$-dimensional manifold, over which the constraints define seven independent Hamiltonian vector fields  which can be integrated into seven-dimensional \emph{gauge orbits}. The seven constraints close the Lie algebra of the similarity group ${\sf Sim}= \mathbbm{R}^+\ltimes ISO(3) $. Except at some singular points which can be ignored because they form a set of measure zero (according to the measure we are going to introduce below), one and only one gauge orbit passes through each point on $\jan$, and therefore the quotient space of $\Sigma$ by the orbits (`reduced phase space') $\Gamma_\st{red} = \jan / \Sim$ is well defined~\cite{HenneauxTeitelboim} and is a symplectic manifold.  In our case we can call this space \emph{shape phase space}.
 
 The Hamiltonian constraint 
 \begin{equation} \label{HamConstraint}
H = \sum_{a=1}^N \frac{\|{\bf p}^a \|^2}{2 m_a} +V_\st{New} = E_\st{cm}, \qquad V_\st{New} = -\sum_{a<b}{m_am_b\over r_{ab}},
\end{equation}
is different: it generates the dynamics of the system and is first-class with respect to ${\bf L}$ and ${\bf P}$. But $H$ and $D$ are second-class, 
\begin{equation}
\{ H , D \} = 2 \sum_{a=1}^N \frac{\|{\bf p}^a \|^2}{2 m_a} + V_\st{New} = 2 \, H - V_\st{New} \neq 0, \label{Lagrange-Jacobi-Relation}
\end{equation}
which is a quantity that never vanishes. This is the phase-space formulation of the Lagrange--Jacobi relation presented above. Therefore each dynamical orbit generated by $H$ (the integral curves of the Hamiltonian vector field associated with $H$) intersects $D=0$ and $\jan$ only once, at the Janus point. In the next section we will use this fact to specify mid-point data at the surface $\jan$, and to identify the space of solutions (and the measure on it) with $\Gamma_\st{red}$. This will not be the end of the story, because the solution space thus defined will still be non-compact. The final step will be to use dynamical similarity (realized, in phase space, by the equivalence relation ${\bf p}_a \sim k \, {\bf p}_a$) to further quotient $\Gamma_\st{red}$ and obtain the space of \emph{physically distinct solutions} $PT^*\shs$ (the notation will be clear in the following), which is compact with respect to a naturally defined measure on it.

\section{\label{typ} Defining Typicality of Microstates in Cosmology}

A review of the conditions that make it possible to define entropy and typicality in standard situations in thermodynamics and statistical mechanics makes it clear that some modification of concepts must be made for the universe. We have seen little discussion of this issue in the literature except for an approach initiated by Gibbons, Hawking and Stewart~\cite{GHS} in an attempt to quantify the typicality of cosmological solutions that lead to a given number of inflationary e-foldings.\,\footnote{\,All of these studies are of the massively symmetry-reduced FLRW models with a homogeneous scalar field. All shape (conformal) degrees of freedom are frozen. In contrast, we aim to develop a theory of the typicality of universes without any symmetry reduction; our only restriction is that the universe is assumed to be spatially closed (and globally hyperbolic).} We follow their basic procedure but with important modifications that enable us to avoid Schiffrin and Wald's criticisms~\cite{SW} of the GHS approach.
\begin{itemize}
\item
 GHS want to define a measure on the space of solutions,  assuming a Hamiltonian system with time-independent Hamiltonian.
\item
Each solution will lie on a constant-energy surface $H(p,q) = E$ on which there is a well-defined induced Liouville measure, so they fix the energy once and for all and define a measure on the solutions with energy $E$. In our case $E=0$.
\item
In order to count each solution once, they assume the existence of a surface that intersects the constraint surface $H=E$ transversely and once only for each solution. In general systems one is able to guarantee these two conditions only locally (in a neighbourhood of each phase-space point). We have a natural choice for this surface: $D=0$, which is a good global gauge-fixing of $H=0$ because it intersects it transversely and once only per solution, as we proved above. This is a very significant advantage not present in \cite{GHS}.
\item
Having obtained their intersection surface, GHS proceed to prove that the 2-form $\omega_{n-1}$ induced on it by the natural symplectic form of phase space $\omega_n = \d p_i \wedge \d q^i$ is invariant under the Hamiltonian phase flow (the Lie derivative with respect to the Hamiltonian vector field associated to $H$). This implies that the associated volume form on the constraint surface $\omega_{n-1}^{\wedge(n-1)}$ (the $(n-1)$-th exterior power of  $\omega_{n-1}$) is invariant too, and consequently it is independent of the choice of intersecting surface.
\end{itemize}

These results can be better expressed with the language of constrained Hamiltonian systems: the measure is \begin{equation}
\mu_{n-1} = | \det \{\chi , H \} | \delta(\chi) \delta(H) \prod_{i,k} \d p_i \d q^k,
\end{equation}
where $| \det \{\chi , H \} |$ is the Faddeev--Popov determinant (in this case, with only a pair of second-order constraints; it is just the absolute value of the Poisson bracket between $H$ and $\chi$). The measure $\mu_{n-1}$ is invariant under the Hamiltonian phase flow.

Moreover if there are further constraints $H_a$, $a=1,\dots,m$, as in our case, one specifies an identical number of gauge-fixings $\chi_a$, and the measure 
\begin{equation}
\mu_{n-m} = | \det \{\chi_a , H_b \} | \prod_{a,b} \delta(\chi_a) \delta(H_b) \prod_{i,k} \d p_i \d q^k
\end{equation}
does not depend on the choice of gauge-fixings.\,\footnote{\,The constraints and gauge-fixings $\varphi_\mu = (H_1 , \dots ,H_m ,\chi_1 , \dots , \chi_m)$, must satisfy some regularity conditions (the Jacobian matrix $\frac{\partial \varphi_\mu}{\partial (p,q)}$ must have rank $2m$, or alternatively the gradients $d H_1, \dots , \d H_m , \d \chi_1, \dots , \d \chi_m$ must be locally linearly independent on the constraint surface, so that  $d H_1 \wedge  \dots \wedge \d H_m  \wedge \d \chi_1 \wedge  \dots \wedge \d \chi_m \neq 0$) (see~\cite{HenneauxTeitelboim} sec. 1.1.2).}

The last remark allows us to improve on the GHS measure. In fact the second-class pair
\begin{equation}
H = \sum_a \frac{{\bf p}^a \cdot {\bf p}^a}{2 m_a} + V_\st{New} \,, \qquad D = \sum_a {\bf r}_a \cdot  {\bf p}^a,
\end{equation}
has the special property that each of its two members taken separately closes two distinct first-class systems, $(H,{\bf P},{\bf L})$ and  $(D,{\bf P},{\bf L})$, with the spatially-relational constraints ${\bf P}$ and ${\bf L}$: 
\begin{align}
&\{ H ,{\bf P}\} = 0,  & &   \{ H ,{\bf L}\} =0,  & &  \{ L^i  , P^j\} = \epsilon^{ijk} P_k,
&\\
&\{ L^i  , P^j\} = \epsilon^{ijk} P_k,  & &  \{ {\bf P} , D\} = {\bf P},  & &   \{ D ,{\bf L}\} = 0. &
\end{align}
This in turn implies that the measure
\begin{equation}
\sigma_H =  | \det \{ (D,\chi_{\bf P} , \chi_{\bf L} ) , (H,{\bf P} , {\bf L} ) \} |  \delta(D) \delta(\chi_{\bf P}) \delta (\chi_{\bf L} ) \delta(H)  \delta({\bf P}) \delta ({\bf L} ) \prod_{i,k} \d p_i \d q^k,
\end{equation}
which is based on gauge-fixing of the system $(H,{\bf P},{\bf L})$ by $D$ ($\chi_{\bf P} $ and $\chi_{\bf L}$ are appropriate gauge-fixings of ${\bf P} $ and ${\bf L}$\,\footnote{\,A very natural choice of $\chi_{\bf P} $ and $\chi_{\bf L}$ is the following: $\chi_{\bf P} = \Rcm = \frac 1 M \sum_a m_a {\bf r}_a$, the coordinates of the centre-of-mass, and $\chi_{\bf L} = ( \mathbbm I_\st{cm}^{12},\mathbbm I_\st{cm}^{13},\mathbbm I_\st{cm}^{31})$, where 
$$
\mathbbm{I}_\st{cm} =\sum_{a=1}^N m_a\left( \mathbbm{1} \, {\bf r}_a^\st{cm}\cdot{\bf r}_a^\st{cm} - {\bf r}_a^\st{cm}\otimes{\bf r}_a^\st{cm} \right),
$$
is the inertia tensor. This last gauge-fixing requires the inertia tensor to be diagonal, so that the axes of the system are aligned with the principal axes of inertia. This gauge-fixing is not always well-defined: in collinear configurations, for example, it fails because the inertia tensor has rank $1$, but these configurations are of measure zero and need not concern us here. The Faddeev--Popov determinant of the second-class system $({\bf P}, {\bf L} ,\chi_{\bf P}, \chi_{\bf L})$ is just (proportional to) $\det \mathbbm I_\st{cm}$, the determinant of the inertia tensor. Then $\chi_{\bf P}$ and $\chi_{\bf L}$ are first-class wrt both $D$ and $H$ (the only nontrivial one is $\{ H , \chi_{\bf L} \}$, which is a linear combination of $D$ and ${\bf L}$).}), is identical to the measure
\begin{equation}
\sigma_D =  | \det \{ (H,\chi_{\bf P} , \chi_{\bf L} ) , (D,{\bf P} , {\bf L} ) \} | \delta(H) \delta(\chi_{\bf P}) \delta (\chi_{\bf L} ) \delta(D)  \delta({\bf P}) \delta ({\bf L} ) \prod_{i,k} \d p_i \d q^k,
\end{equation}
based on gauge-fixing of the system $(D,{\bf P},{\bf L})$ by $H$. 
In fact, on the constraint surface ${\bf P} = {\bf L} =0$ we can write the first Faddeev--Popov matrix as
\begin{equation}
\left(
\begin{array}{ccc}
 \{H,D\} & \left\{H,\chi _{\bf P}\right\} & \left\{H,\chi _{\bf L}\right\} \\
 \{ {\bf P},D\} & \left\{ {\bf P},\chi _{\bf P}\right\} & \left\{ {\bf P},\chi _{\bf L}\right\}  \\
 \{ {\bf L},D\} & \left\{�{\bf L},\chi _{\bf P}\right\} & \left\{ {\bf L},\chi _{\bf L}\right\}   \\
\end{array}
\right)
\approx
\left(
\begin{array}{ccc}
 \{H,D\} & \left\{H,\chi _{\bf P}\right\} & \left\{H,\chi _{\bf L}\right\} \\
{\bf 0} & \left\{ {\bf P},\chi _{\bf P}\right\} & \left\{ {\bf P},\chi _{\bf L}\right\}  \\
{\bf 0} & \left\{�{\bf L},\chi _{\bf P}\right\} & \left\{ {\bf L},\chi _{\bf L}\right\}   \\
\end{array}
\right)
\end{equation}
because of course $ \{ {\bf P},H\}  \approx   \{ {\bf L},H\} \approx {\bf 0}$,
and the other Faddeev--Popov matrix as
\begin{equation}
 \left(
\begin{array}{ccc}
 \{D , H \} & \left\{D,\chi _{\bf P}\right\} & \left\{D,\chi _{\bf L}\right\} \\
 \{ {\bf P},H\} & \left\{ {\bf P},\chi _{\bf P}\right\} & \left\{ {\bf P},\chi _{\bf L}\right\}  \\
 \{ {\bf L},H\} & \left\{{\bf L},\chi _{\bf P}\right\} & \left\{ {\bf L},\chi _{\bf L}\right\}   \\
\end{array}
\right)
=
 \left(
\begin{array}{ccc}
 \{D , H \} & \left\{D,\chi _{\bf P}\right\} & \left\{D,\chi _{\bf L}\right\} \\
{\bf 0} & \left\{ {\bf P},\chi _{\bf P}\right\} & \left\{ {\bf P},\chi _{\bf L}\right\}  \\
{\bf 0}& \left\{{\bf L},\chi _{\bf P}\right\} & \left\{ {\bf L},\chi _{\bf L}\right\}   \\
\end{array}
\right)
\end{equation}
(in this case because $ \{ {\bf P},D\}  \approx   \{ {\bf L},D\} \approx {\bf 0}$), and their determinants
will coincide in absolute value:
\begin{equation} \!\!\!\!\!\!
\det   \left(
\begin{array}{ccc}
 \{D , H \} & \left\{D,\chi _{\bf P}\right\} & \left\{D,\chi _{\bf L}\right\} \\
{\bf 0} & \left\{ {\bf P},\chi _{\bf P}\right\} & \left\{ {\bf P},\chi _{\bf L}\right\}  \\
{\bf 0}& \left\{�{\bf L},\chi _{\bf P}\right\} & \left\{ {\bf L},\chi _{\bf L}\right\}   \\
\end{array}
\right)
=
- \det \left(
\begin{array}{ccc}
 \{H,D\} & \left\{H,\chi _{\bf P}\right\} & \left\{H,\chi _{\bf L}\right\} \\
{\bf 0} & \left\{ {\bf P},\chi _{\bf P}\right\} & \left\{ {\bf P},\chi _{\bf L}\right\}  \\
{\bf 0} & \left\{{\bf L},\chi _{\bf P}\right\} & \left\{ {\bf L},\chi _{\bf L}\right\}   \\
\end{array}
\right).
\end{equation}
We see that the fact that the two constraints $D$ and $H$ separately close first-class systems
with the gauge constraints ${\bf P}$ and ${\bf L}$ is crucial for the equivalence of the two measures.
This property is what lies behind the mechanism called \emph{symmetry trading}, which is at the
basis of Shape Dynamics \cite{GGK,FlaviosSDtutorial}, and we see it here realized. It gives us a further significant advantage compared with \cite{GHS}: direct access to a gauge-invariant measure.

The system $(H,{\bf P},{\bf L})$ now describes the solutions of the $E={\bf L}=0$ Newtonian $N$-body problem
in a temporally relational fashion,\,\footnote{\,The Hamiltonian being a constraint embodies a description that does not depend on any particular time parametrization. The system $(H,{\bf P},{\bf L})$ realizes \emph{spatial and temporal relationalism}  \cite{FlaviosSDtutorial}.} and the gauge orbits of $(H,{\bf P},{\bf L})$ on the constraint surface $H={\bf P}={\bf L}=0$ are solutions of the relational \nbn. On the other hand, the system $(D,{\bf P},{\bf L})$ is a purely kinematical description of $N$-body \emph{shapes} and their conjugate momenta;\,\footnote{\,The system $(D,{\bf P},{\bf L})$ in its turn embodies \emph{scale and spatial relationalism}   \cite{FlaviosSDtutorial}.} it knows nothing of dynamics and solution spaces. Following Arnold \cite{arnold1989mathematical} (Appendix~5), we notice that the system $(D,{\bf P},{\bf L})$  
realizes a Poisson action of the Lie group $\Sim$ on the $N$-body phase space $\Gamma$. Then there is a unique construction called \emph{reduced phase space} which gives a symplectic manifold $\Gamma_\st{red}$ obtained by quotienting the constraint surface $D={\bf P}={\bf L}=0$ by the gauge orbits. Our measure $\sigma_D$ is just the volume form associated with the symplectic form of $\Gamma_\st{red}$. Moreover,  Arnold proves (a slightly more general version of) the following theorem:
\begin{quotation}\noindent
\it The reduced phase space $\Gamma_\st{red}$ is symplectic and diffeomorphic to the cotangent bundle of the quotient configuration manifold $\shs$ (shape space).
\end{quotation}
In particular, the canonical symplectic form of the cotangent bundle is equivalent (symplectic) to that of reduced phase space, and therefore the volume form $\sigma_D$ is given by the canonical one on $T^* \shs$. This construction is completely natural and unique.

To summarize, we have proven the equivalence of our GHS-type measure on the space of solutions of the $E ={\bf L} =0$ $N$-body problem to the measure on the reduced phase space $\Gamma_\st{red}$ of the scale- and spatially-relational system $(D,{\bf P},{\bf L})$, which we call \emph{shape phase space}. Moreover, this measure is the same as the canonical volume form on $T^*\shs$, the cotangent bundle to shape space.

But we still do not have exactly what we need: the measure on $T^*\shs$ is infinite. In fact one can easily see that shape space $\shs$ is compact (gauge-fixing $D$ by fixing the moment of inertia results in a hypersphere), but its cotangent bundle is not compact according to the canonical measure: one can take arbitrarily large cotangent vectors. However, as we pointed out above, there is a further redundancy: rescaling the shape momenta just gives a similar solution (meaning rescaled by the same amount at all times). It is then clear that, if we mod out by momentum rescalings (and reflections, because changing the sign of the mid-point momenta gives the same solution, just pointing towards the opposite time direction),
we will end up with a compact space. We can call this the solution space (or the space of physically distinct solutions).  It coincides with $PT^*\shs$, the \e{projectized cotangent bundle} to shape space. Thus, although the relational \nb is not fully scale-invariant, we have a complete \e{non-redundant scale-invariant representation} of its solution space.


However, $PT^*\shs$ is not a symplectic manifold, because it is odd-dimensional. As the final step in establishing the mathematical structures we need, we would like to define a measure on it, but the measure on $T^*\shs$ does not project uniquely to a measure on $PT^*\shs$. Some more input is needed. Here there must be a choice on our part, because the natural geometrical structures possessed by the various configuration spaces we considered so far do not suffice to specify uniquely a measure on $PT^*\shs$. One way we could induce a measure on $PT^*\shs$ is to fix the norm of cotangent vectors to $1$. This gives the \emph{sphere bundle} $T_1^*\shs$~\cite{blair2010riemannian}, which is the double cover of $PT^*\shs$ (each point in $PT^*\shs$ maps to two points related by momentum reflection in $T_1^*\shs$). But in order to have a norm on cotangent vectors we need a metric on $\shs$, and this is where our choice enters.

The extended configuration space ${\sf Q}$ is endowed with a natural metric, the \emph{mass metric} \cite{FlaviosSDtutorial},
\begin{equation}\label{MassMetricOnS}
\d s^2 = \sum_{a=1}^N m_a \, \d {\bf r}_a \cdot \d {\bf r}_a,
\end{equation}
and, as shown by Montgomery~\cite{Mont} (p. 320), this induces a metric on the quotient of ${\sf Q}$ by the isometry group of $\d s^2$ (a construction called \emph{Riemannian quotient}). But the metric above is not invariant under dilatations, and therefore the Riemannian quotient procedure does not apply to $\shs = {\sf Q}/{\sf Sim}$. We need a metric whose isometry group is ${\sf Sim}$. Any metric that is conformally related to (\ref{MassMetricOnS}) through a factor that is homogeneous of degree $-2$ in ${\bf r}_a$ would suffice. In this paper we appeal to the principle of simplicity and make the choice
\begin{equation}
\d s^2= \sum_{a=1}^N m_a  \frac{\d {\bf r}_a \cdot \d {\bf r}_a}{I_\st{cm}},\label{simple}
\end{equation}
which is the simplest we can imagine. Any other choice would seem to lack a sufficient reason. Ultimately, our choice, for all its simplicity, must (like any other choice) be tested through the predictions it makes for observable quantities. We will discuss this below.

The metric (\ref{simple}) gives the following norm for a cotangent vector ${\bf p}_a$:
\begin{equation}
I_\st{cm} \sum_{a=1}^N \frac{{\bf p}^a \cdot  {\bf p}^a}{m_a},
\end{equation}
which we may set equal to one, obtaining a measure on $T_1^*\shs$ and consequently on $PT^*\shs$. This measure is finite and probabilities are well-defined with it.

For illustration, in Appendix A we flesh out the 3-body case in all its details, showing explicitly the analytic form of the measure on $PT^*\shs$.

\subsection*{How we avoid Schiffrin and Wald's criticisms}

In the remainder of this section, we indicate how we avoid the two most serious difficulties identified by Schriffrin and Wald~\cite{SW} in the approach initiated by Gibbons, Hawking and Stewart~\cite{GHS}.

The first of these relates to the conditions under which a count of microstates is physically justified and well defined. If the microstates are discrete and finite in number, the count of them is unproblematic. However, if there is a continuum of microstates, the count must be based on a measure, which must be finite if unambiguous statements are to be made. This is an essential condition. Schiffrin and Wald point out that the various studies based on the approach of~\cite{GHS} gave discordant results because they used an infinite measure, in which case probabilities become hopelessly ambiguous.\,\footnote{\,The probability of a finite-measure region would always be zero. That of a region whose complement has finite measure would always be 1, and an infinite-measure region whose complement has infinite measure too would have an indefinite probability.} At least in the \nbn, we resolve this problem because we work on the solution space $PT^*\shs$, which has a finite number of degrees of freedom and, as we have just shown, can be ascribed a finite measure.\,\footnote{\,There are two aspects of our measure that require a good motivation. One, our choice of (\ref{simple}), we have already discussed. The other is our use of the Liouville measure induced on $T^*\shs$. This appears as a natural, indeed obvious choice. The very possibility of defining Newtonian dynamics of point particles interacting through universal gravitation derives ultimately from the geometrical properties of Euclidean space and the notion of mass. Taken together in the form of the mass-weighted coordinates $\sqrt{m_a}\,{\bf r}_a$, they lead directly to the kinetic metric $\sum_a\,m_a\,\d {\bf r}_a\cdot \d {\bf r}_a$ that underlies Newtonian dynamics. The rules for variation of the action lead to the distinguished status of the canonical momenta and the definition of the symplectic form. Thus, the Liouville measure directly reflects the fundamental structure of phase space, which (as we have noted) itself derives from geometry and the notion of mass.},\,\footnote{\,\label{bounce}In loop quantum cosmology a `bounce' replaces the big-bang singularity of classical general relativity. Ashtekar and Sloan~\cite{AshtekarSloan} exploit the measure induced at the bounce to specify initial conditions at the `bounce'. This enables them to define a measure on the space of solutions of symmetry-reduced FLRW solutions with a homogeneous scalar field and estimate probabilities for the number of inflationary e-foldings. The bounce is clearly a quantum Janus point. In contrast to \cite{AshtekarSloan}, we restrict ourselves in this paper to purely classical theories with a Janus point and do not make any symmetry reduction, which in the case of the \nb would correspond to consideration of only those solutions in which the shape of the system does not change (homothetic motions~\cite{GibbonsEllis}). We thank Sean Gryb for drawing our attention to \cite{AshtekarSloan}} In the field-theoretic context of general relativity, in which the classical theory has infinitely many degrees of freedom, it may be necessary to assume \cite{SW} that, in some as yet unknown way, quantum gravity supplies an effective cutoff.


The final conceptual issue that we must discuss concerns our basic statistical assumption concerning the probability with which particular solutions are likely to be realized. The situation is this. We assume we know the law of the universe and that, as in the \nb with $E={\bf L}=0$, all of its solutions have a Janus point. Then each point on $PT^*\shs$ stands in a one-to-one relation with one of the distinct solution curves in $\shs$. We know the irreducible miminum of data needed to fix a solution. What we do not know is the probability with which any particular solution will be realized. The law by itself gives no guidance on this matter.

Under these circumstances there is no rational stance we can adopt other than Laplace's \e{principle of indifference}:\,\footnote{\,Laplace presented the principle without a name as more or less obvious. The coining `principle of indifference' is due to the economist John Maynard Keynes in his \emph{Treatise on Probability} (1921).} a rational agent who knows a certain number of situations can be realized but has no other information can do no better than ascribe equal probabilities to the various outcomes. This was clearly good advice for any gamblers whom Laplace might have advised on betting tactics. In fact, in Bayesian data analysis the principle of indifference is widely used to eliminate bias.

We are, de facto, adopting \e{the fundamental postulate} of statistical mechanics, according to which the occupation of any  state is assumed to be equally probable to that of any other state\id $p_i = 1/\Omega$, where $\Omega$ is the number of accessible states when their number is finite. One is usually justified in making this assumption for an isolated system for which one has good reason to believe that it is in equilibrium. One assumes that the system is ergodic, so that averages over time can be replaced by averages over phase space. It is obvious that we are applying the postulate under entirely different circumstances.

Indeed, one might think that nothing can be predicted on the basis of ignorance.  However we will show that, applied appropriately,  it has the potential to lead to \e{strong predictions for the mid-point data}, the consequences of which can in principle be tested observationally. We say `in principle' because the actual calculations needed to arrive at  definite predictions for observable quantities are subject to uncertainties, above all ones related to the matter content of the universe. However, what we say about prediction of mid-point data is in effect closely analogous to one of the most important insights in statistical mechanics. Suppose we have a dynamical system that has a known fixed energy and consists of a gas of a very large number of `elastic balls' in a box of fixed size with walls off which the balls bounce elastically. Then the overwhelming majority of microstates, coarse grained and counted using the Liouville measure, correspond to a very good approximation to a Maxwell--Boltzmann distribution. If one now makes the Laplace assumption that all microstates are equally probable, this corresponds to an extremely strong prediction of what will be found on a \e{single} opening of the box.

That one does in the overwhelming majority of cases find a Maxwell--Boltzmann distribution is explained by the equilibration processes known (and very largely understood) to take place in such a box in which the system is initially in disequilibrium. An immense body of experimental data coupled with dynamical modelling therefore lies behind the successful prediction just described. Our situation is quite different: the \e{single} opening of the box corresponds to training our telescopes on the sky to see what our universe is like. But this is a one-off test. No cosmologist has made repeated observations of universes evolving, let alone equilibrating.

Thus, we have no accumulated empirical justification for the principle of indifference. It is rather a case of necessity being the mother of invention. It forces us to make a clean and simple conjecture of the kind that Popper advocated \cite{Popper}. The \e{sine qua non} for us to be able to do this is mathematical consistency. This we have established through our demonstration that there does exist a well-motivated finite measure for the universe, treated as a scale-invariant system of evolving shapes.

If the nature of our statistical assumption matches standard practice, the use we hope to make of it is very different. The fundamental postulate of statistical mechanics tells us what to expect at the end of a process: an equilibated state of heat death. Our use of the principle tells us what to expect on the Janus surface $\jan$ and thus provides the initial data for the evolution in both directions away from $\jan$. It does not tell us how things end but how they \e{begin}. 

\section{Choice of the Primary State Function}

Our system is characterized by a number of state functions (functions only of the current state of the system which do not depend on \emph{how} the system got there). There can be up to $6N-14$ such independent functions. In the first instance, we can consider the universe as a simple system in 
the sense of Lieb  and Yngvason~\cite{lieb2004guide}, that is, a system that is characterized by a `primary' state function (typically the energy in the examples given in~\cite{lieb2004guide}). Our primary state function, however, cannot be the energy, because it is fixed to zero. Moreover the energy is a dimensionful quantity, and our whole approach is based on considering as physically distinct only solutions that differ in shape space. This implies that our primary state function has to be dimensionless (scale-invariant). It is then natural to take the \emph{shape complexity function} we identified in~\cite{BKM3} as our primary state function:
\bq
C_\st{S}= \frac{\ell_\st{rms}}{\ell\st{mhl}}\equiv {1\over M^{5/2}}I_\st{cm}^{\,1/2}V_\st{New},~~V_\st{New}=\sum_{a<b}\,{m_am_b\over r_{ab}},\label{comp}
\ee
where $\ell_\st{rms}$ was defined in (\ref{rms}).

In~\cite{BKM3}, in fact, we observed that $C_\st{S}$ is a scale-invariant property of the system that characterizes its state rather well: it expresses how clustered/inhomogeneous its instantaneous shape is.
In~\cite{BKM3} we proved that $C_\st{S}$ is bound to grow on average away from the Janus point (although it fluctuates, one can bound $C_\st{S}$ from below and above with linear functions of Newtonian time). We also made the observation that  $C_\st{S}$ measures well the most striking characteristic of the $N$-body solutions, that is, the fact that the system becomes more clustered as it evolves away from $\janp$, and therefore $C_\st{S}$ would be the simplest way to describe qualitatively the difference between states of a solution near to and far from its Janus point. In this paper we make this quantitative and precise.

In fact, in the first place we use $C_\st{S}$ to characterize the typicality of mid-point data on $\jan$. We believe it is the natural, uniquely suitable counterpart of the energy in conventional statistical mechanics. For the inherently nonequilibrium and unconfined relational universe, there are no readily available counterparts of the `work functions' (pressure, volume and temperature), which played such an important role in creating the framework in which thermodynamics was discovered. It is however noteworthy that the relational principles that underlie our approach dictate quotienting by scale, which, together with dynamical similarity, leads directly to the finite induced measure on $PT^*\shs$. This is the condition that in standard statistical thermodynamics is achieved by restriction of the system to a finite volume. The relational universe `comes with its own box'. 

This key fact, together with the existence of the primary state function, means that an adequate statistical framework does exist in the cosmological context. When we discussed Laplace's principle of indifference, we commented that a known law gives no guidance which solution of it will be realized and, for this reason, there appeared to be no alternative to ascribing equal probability of occurrence to each of them. It is therefore very striking that, given our need to employ Laplace's principle if we are to establish the typicality of solutions, the relational \nb gives us precisely the structures we need to do that: the Janus surface $\jan$, whose quotient by the gauge orbits of $\Sim$ is symplectic to $T^*\shs$; the space of physically distinct solutions, which coincides with $PT^*\shs$; the Liouville measure on $T^*\shs$ and a metric on $\shs$ which allow us to induce a finite measure  on $PT^*\shs$; finally, a uniquely defined primary state function.

Further state functions can be used to achieve a more restrictive definition of classes of microstates considered to be macroscopically indistinguishable. Complexity presents itself as the most immediately obvious choice by its simplicity and close analogy with energy.\,\footnote{In fact, as we noted in \cite{BKM2,BKM3}, the negative of the complexity is the \e{shape potential}\id the part of the Newton potential that changes the shape of the universe but not its size. This is exactly what one requires of a theory in which overall scale is gauge. Moreover, since the total energy of the relational \nb is zero, specification of the complexity, and thus the shape potential, is closely analogous to specifying the energy in conventional statistical mechanics.} There are many `finer' measures of macroscopic distinguishability, e.g., the complexity of subsystems or the power spectrum of density fluctuations. In principle, one could introduce a complete set of independent state functions, at which point all `entropic ignorance' is eliminated and one is dealing with individual microstates~\cite{Jaynes-BoltzmannVsGibbs}. What one needs are a few state functions that capture the most salient features of the evolving microstates. In this paper, we shall merely show what can be done with the primary state function: it is adequate to study the typicality of solutions and an excellent first diagnostic of the effect of evolution off the Janus surface.

\label{SectionPrimaryStateFunction}

\section{Typicality of Solutions and Microstates}

The most important fact that we have so far established is this: every point in $PT^*\shs$ uniquely and non-redundantly defines a solution. Using our measure  on $PT^*\shs$, we can `count' the solutions whose solution-determining mid-point $\janp$ have given values $c$ of the primary state function $C_\st{S}$.
Consider the natural projection $\pi$ of points $\janp$ in $\jan$  to  $PT^*\shs$ (this projection associates each point on the surface $\jan$ with the corresponding gauge orbit in $T^*\shs$, and then projects each cotangent vector in $T^*\shs$ onto the ray in $PT^*\shs$ that is parallel to it). Then we define
\bq
{\mathcal E}_\st{sol}(c)=\textrm{Vol} \left\{\pi(\janp) \in  PT^*\shs ~/~ C_\st{S}(\janp) = c \right\}, \label{solent}
\ee
where the volume $\textrm{Vol} $ is calculated through the measure on $PT^*\shs$ we defined in the previous section.
As a count of states with given value of a phase-space function, ${\mathcal E}_\st{sol}$ is clearly an entropy-type quantity. To avoid confusion, we call it the \e{solution entaxy}: each $\janp$ determines a solution (sol); `taxis' is Greek for order and `en' means `towards'. As we shall see, ${\mathcal E}_\st{sol}$ is a measure of creation by gravity of order out of disorder.

We use the simple volume rather than a $p_i\,\textrm{log}\,\,p_i$ integral because we assume each possible solution is equally likely to be realized. This is the fundamental postulate of statistical mechanics (Laplace's principle of indifference) applied to solutions rather than states. It is also an example of the maximum-entropy principle used to avoid bias~\cite{Jaynes-BoltzmannVsGibbs} in a situation in which one has only partial information. In such a case, one \e{maximizes} the relevant $p_i\,\textrm{log}\,\,p_i$ expression subject to constraints supplied by known prior facts. In our case there is no prior knowledge -- given a law, one has no information which of its solutions will be realized. It is well known that in  the absence of prior information the $p_i\,\textrm{log}\,\,p_i$ expression is maximized when all the probablities $p_i$ are equal. The $p_i\,\textrm{log}\,\,p_i$ expression then reduces to the logarithm of the measure. It is important to retain the logarithm in the statistical-mechanical definition of entropy because then the entropy of independent systems is additive as is found empirically in phenomenological thermodynamics. However, in our case we are considering a unique system -- the universe -- so the need for the logarithm falls away.

Our other important assumption is the measure. Since the whole theory is expressed through a phase space with a dynamically distinguished Liouville measure, it  seems unnatural to choose any other measure. However, the ultimate justification of both our assumptions, which are questioned in~\cite{SW}, must be a posteriori through observations, which we discuss below. 
Note, however, that our approach is not subject to the most important objection raised in~\cite{SW} against typicality arguments of the kind we adopt: our measure is finite, not infinite, so our probability estimates are unambigous.\,\footnote{\,An analogous argument based on the use of a Janus point, or `bounce', is made in~\cite{AshtekarSloan}. However, whereas `bounce' is clearly an appropriate word to use in the context of loop quantum cosmology, in which a quantum bounce `intervenes' to halt and reverse the collapse inherent in classical general relativity, all of the Janus points we consider correspond to smooth unhindered passage through points whose existence is mandated by the underlying time-symmetric classical equations of motion.}

\begin{figure}[t]
\center\includegraphics[width=0.7\textwidth]{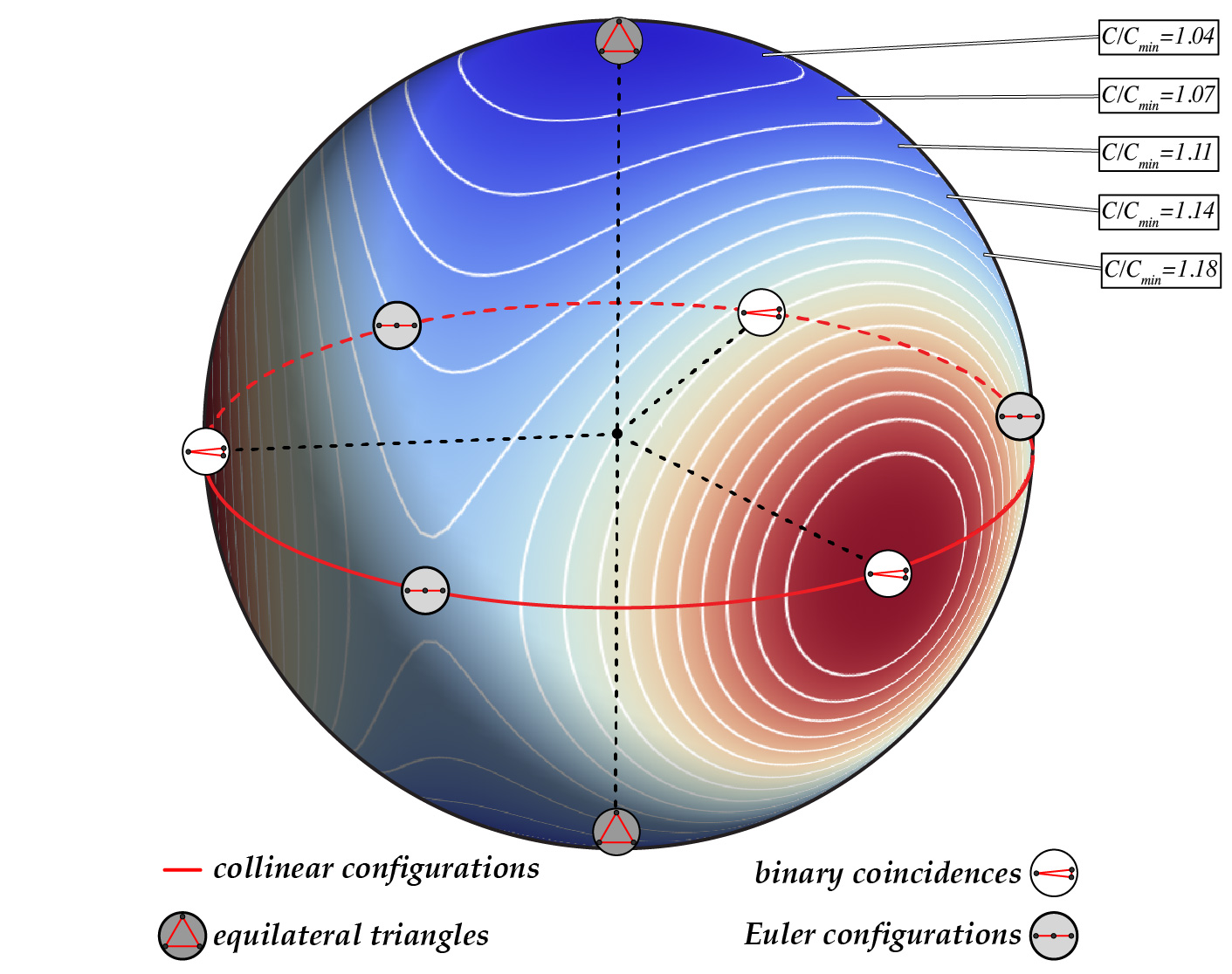}
\caption{\small \small Shape complexity (\ref{comp}) as elevation plot over the equal-mass 3-body shape space. The minimum is at the most `uniform' shape (the equilateral triangle) and grows unboundedly in `cluster formations' ($r_{ab} \ll r_{ac}, r_{bc}$). The lengths of the equal-$C_\st{S}$ curves measure the typicality of the corresponding shapes. For large $N$, minimal $C_\st{S}$ correspond to extremely uniform (super-Poissonian) distributions~\cite{BattyeGibbons}, large $C_\st{S}$ to strongly clustered distributions.
\normalsize} \label{ShapeSpace}
\end{figure}

\begin{figure}
\center\includegraphics[width=0.7\textwidth]{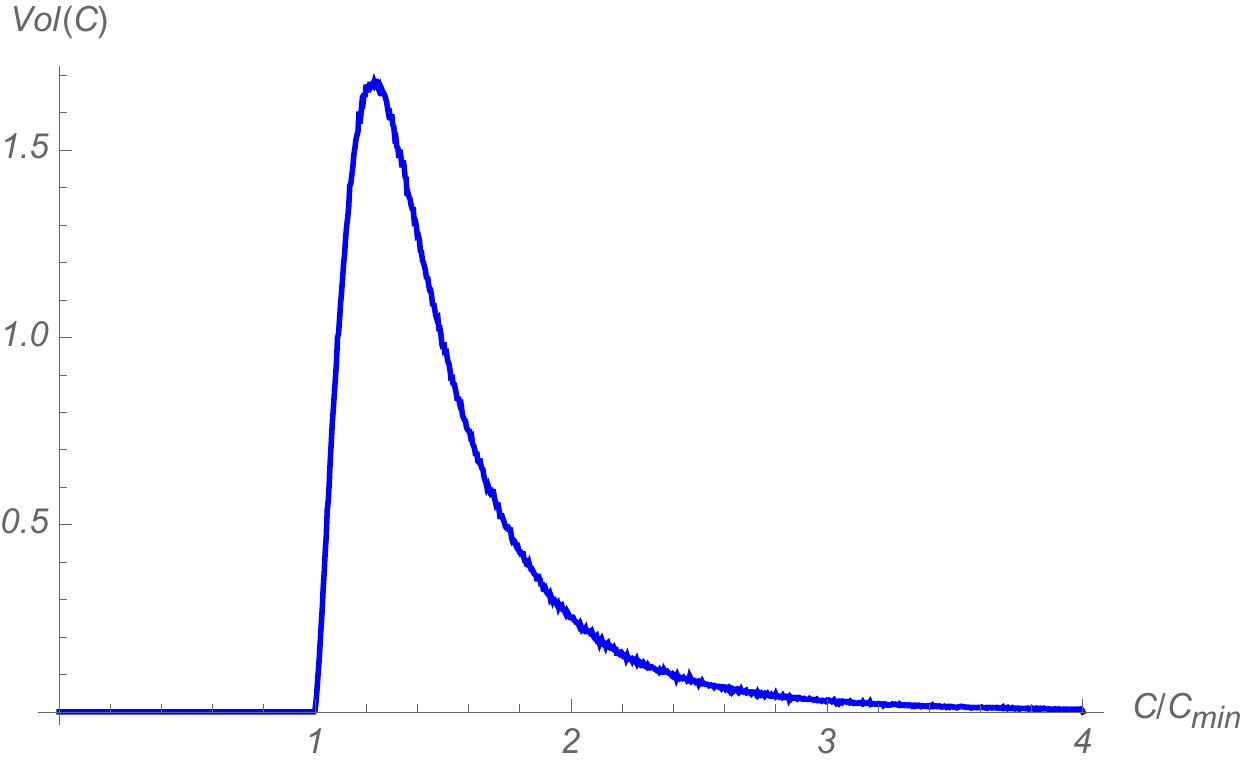}
\caption{\small \small The fraction of the volume of the equal-mass 3-body shape space occupied by 
configurations with different values of the complexity $C$. The above diagram should
be understood in the sense of a probability density function: taking an interval $C\in (C_1,C_2)$,
the area below the curve ${\rm Vol}\, (C)$ between $C_1$ and $C_2$ gives the fraction of the volume of $\shs$ 
occupied by states with $C$ between $C_1$ and $C_2$. This diagram has been obtained numerically,
by sprinkling points on $\shs$ with a probability given by the natural measure induced by ${\sf Q}$ on $\shs$. Notice that the curve is, of course, zero for $C \in (0,C_\st{min})$. Then it jumps very quickly to a high value followed by an initially steep but then much more gradual decline. 
\label{Hist3}}
\end{figure}

\begin{figure}
\center\includegraphics[width=0.9\textwidth]{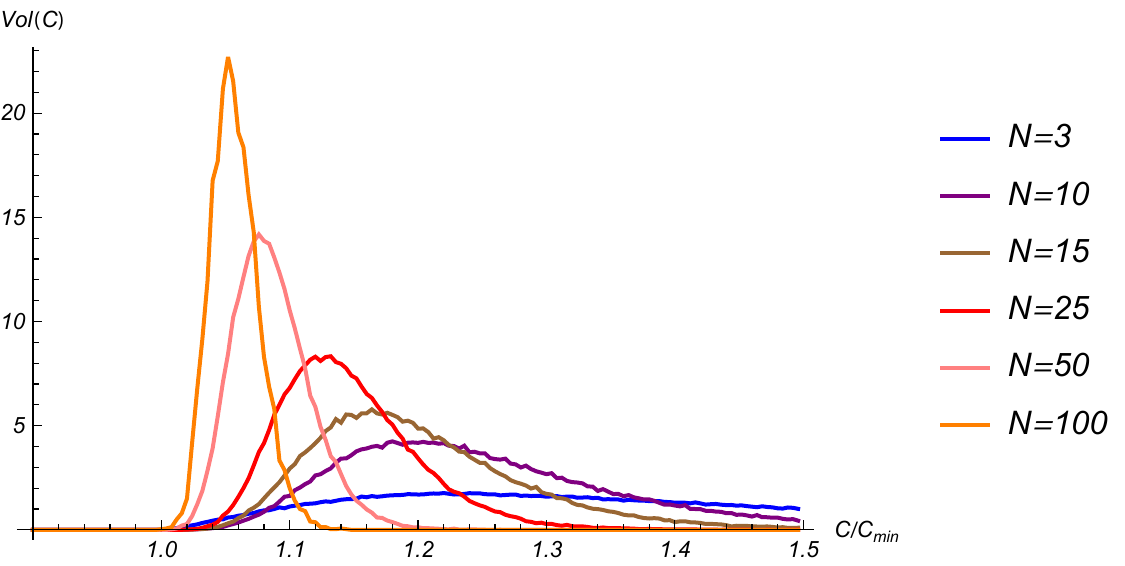}
\caption{\small The fraction of the volume of the $N$-body shape space occupied by 
configurations with different values of $C/C_\st{min}$ for five values of $N$ and for a Gaussian distribution of equal-mass particles. With this normalization,
it is apparent how the distribution gets peaked around smaller values of $C/C_\st{min}$ as $N$ grows,
and how the peak becomes more pronounced.} \label{HistN}
\end{figure}

We now consider a very important relationship, namely the dependence of the solution entaxy ${\mathcal E}_\st{sol}$ on the value of the primary state function $C_\st{S}$ at the Janus point. In accordance with our adoption of Laplace's principle and the induced Liouville measure, every shape is equally likely at $\jan$. Figure \ref{ShapeSpace} shows clearly that the area on the 3-body shape sphere corresponding to shapes near dart-like triangles (with two particles relatively much closer to each other than to the third) is small. In contrast, the moderately scalene triangles occupy the bulk of the shape-sphere area.  Moreover, most of these triangles have $C_\st{S}$ values not much higher than the minimum at the equilateral triangle.
This fact is clearly seen in the histogram in Fig.~\ref{Hist3}. The effect becomes more and more pronounced as $N$ gets larger (Fig.~\ref{HistN}).

Thus, on the the basis of the two assumptions we have made, it follows that the overwhelming majority of solutions will have values of their primary state function $C_\st{S}$ very near but not exactly at its absolute minimum. Thus, the most typical states of the solutions at their respective Janus points will have very uniform spatial distributions of the particles.

The overwhelming majority of solutions will go from $C_\st{S}$ values very close to (but not exactly at) the minimum at  $\jan$ to infinite limits as $t\rightarrow\pm\infty$. In the continuum-approximating limit $N\rightarrow\infty$, the disorder-to-order effect will be very strong.
 
\begin{figure*}[b!]
\includegraphics[width=\textwidth]{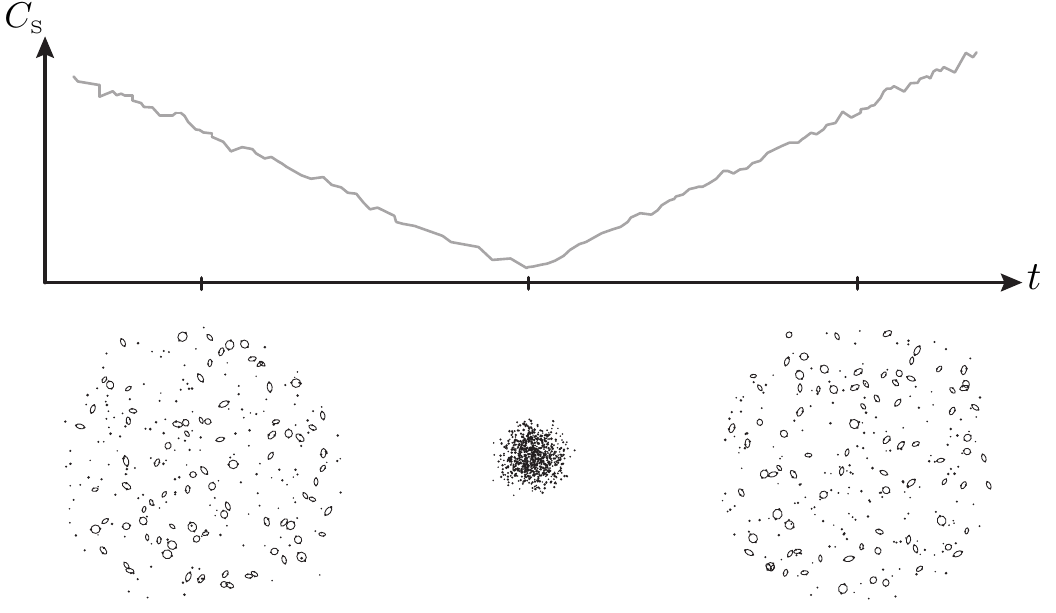}
\caption{\small Top: plot of the shape complexity $C_\st{S}$ found by numerical simulation for $N=1000$ equal-mass particles with Gaussian initial data satisfying $E_\st{cm}={\bf L}={\bf P}=0$. Bottom: `artistic impression'\id not found by numerical simulation, of three (scaled) configurations at different Newtonian times $t$ corresponding to three points in the solution at the top.} 
\label{TriplePictureWithComplexity}
\end{figure*}

We now explain why ${\mathcal E}_\st{sol}$ measures creation of order out of disorder. For this
we must first define the \emph{current} (instantaneous) entaxy of a solution. So far, we have only defined the solution entaxy ${\mathcal E}_\st{sol}$.
At each point $s$ on a solution, the current entaxy ${\mathcal E}(s)$ is defined simply as the entaxy corresponding to the shape $s$:
\bq
{\mathcal E}(s)=\textrm{Vol} \left\{\pi(\janp) \in PT^*\shs ~/~ C_\st{S}(\janp) = C_\st{S}(s) \right\}. \label{solent}
\ee
We noted earlier that solution-determining data away from the Janus point include not only the instantaneous shape $s$ and direction $d$ in $\shs$ but also some quantity that measures the `bending' of the solution curve in $\shs$ at the considered instant. This implies that, for each shape $s$ and direction $d$ in $\shs$, there exists a one-parameter family of solutions, and just one solution of the family has its Janus point at $s$. Our definition of the current entaxy ${\mathcal E}(s)$ corresponds therefore to the ${\mathcal E}_\st{sol}$ of the solution that has its Janus point at the shape $s$.

We showed in~\cite{BKM3} that in every $N$-body solution with $E={\bf L}=0$ the shape complexity $C_\st{S}$ fluctuates around its minimal value in the neighbourhood of the Janus point (where $D=0$) and then grows with fluctuations but within monotonically rising bounds in both directions away from $\jan$ as $t\rightarrow\pm\infty$ (see Fig.~\ref{TriplePictureWithComplexity}). This disorder-to-order effect is already present in the \tbn, in which more or less chaotic 3-body dynamics near $\jan$ evolves in both time directions into non-chaotic motion of an inertial particle and a Kepler pair moving in the opposite direction to the single particle. For $N>3$, the system breaks up into subsystems whose centres of mass separate in Hubble-type motion as in the inertial system considered in Sec.~\ref{SSE}. In the subsystems, energy and angular momentum are increasingly well defined and conserved. Overall, the complete $N$-body system becomes increasingly well ordered. Using the emergent Kepler pairs as rods and clocks, internal observers either side of $\jan$ will find themselves in an expanding universe, as illustrated by Fig.~\ref{TriplePictureWithComplexity}. For details, see \cite{BKM2,BKM3}.

Thus in every solution the primary state function $C_\st{S}$ has minimal values near $\jan$ and then fluctuates between monotonically rising bounds as $t\rightarrow\pm\infty$ (Fig.~\ref{TriplePictureWithComplexity}). Since low $C_\st{S}$ values correspond to unstructured uniformly random states and large values to well separated subsystems with well-defined conserved quantities, it is clear that, with increasing distance from $\jan$, gravity inevitably -- in all solutions -- creates order (and with it an arrow of time) out of random disorder. Our new theory using statistics of mid-point data now suggests we can predict how pronounced the disorder-to-order effect will be.

In Sec.~\ref{SSE}, we questioned the possibility of applying standard entropic arguments to a universe that is unconfined and therefore can expand freely. This led us to cast doubt on the claim that the entropy of the universe is increasing. We have now shown that, at least in the \nbn, entaxy is a sensible substitute for entropy and that it \e{decreases} in all solutions away from the Janus point. Any observer must be on one or the other side of the Janus point and find herself in a universe that is expanding in the direction of decreasing entaxy. In Sec.~\ref{eesl}, we will consider how this conclusion can be reconciled with the second law of thermodynamics. Before that, we will show how one can extract meaningful predictions from our assumptions and some statistical considerations.

\subsection*{Our theory is predictive}\label{OurTheoryIsPredictive}

\begin{figure}[b!]
     \begin{center}
\raisebox{12pt}{\includegraphics[width=0.5\textwidth]{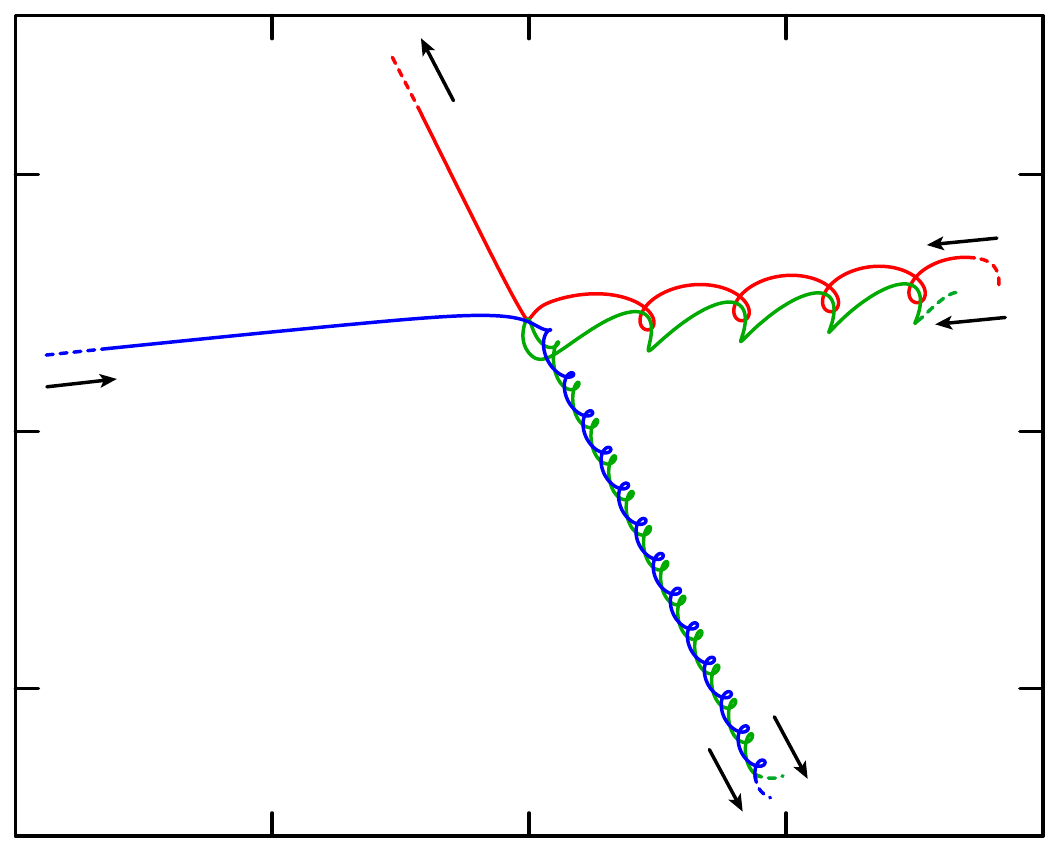}}~~~~~~\includegraphics[width=0.45\textwidth]{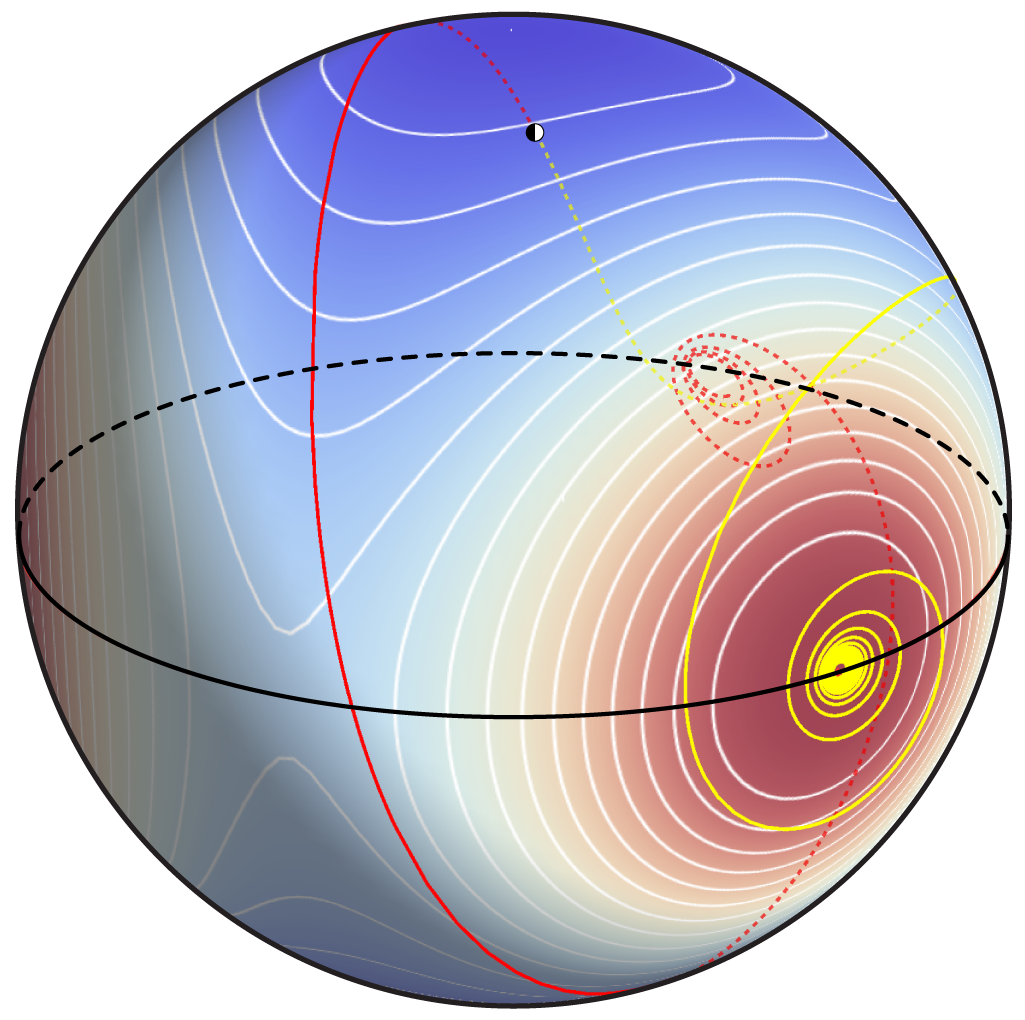}
     \end{center}
\caption{\small Left: A typical solution of the 3-body problem in a centre-of-mass inertial frame in the case $E={\bf L}={\bf P}=0$ (when the motion is always planar). With the time direction show by the arrows, one particle of the pair coming from the right forms a new pair with the particle coming from the left. The Janus point is in the middle of the region of non-trivial 3-body interaction. Right: the same solution plotted  on shape space $\shs$. The red and the yellow part of the orbit are distinguished by belonging, respectively, to the (nominal) past and the 
future of the point at which the dilatational momentum $D$ is zero and the centre-of-mass moment of inertia $I_\st{cm}$ is at its  minimum. This point is shown as $\LEFTcircle$ at the back of $\shs$, where the orbit is dashed.  Whereas the complexity level-lines are shown for the spiralling yellow half, this cannot be done for analogous the red (dotted) half at the `back' of the shape sphere. In this diagram, we see clearly the dramatic effect of the dynamical attractors that act on the shape degrees of freedom when the `veiling' effect of scale is factored out of the conserved Liouville measure on the Newtonian phase space. As with inertial motion, the evolution curves are `sucked' into ever-decreasing regions of shape space. However, at the same time, structure is created (in the form of a Kepler pair in this simplest example of the 3-body problem). Thus, already in the toy universe we also see that Hubble-type expansion is an inevitable attractor in a gravitational theory formulated on an extended phase space with scale as a degree of freedom. Moreover, the attractor effect is accompanied by ever more perfect formation of structure. A similar diagram, for a different solution, can be seen in \cite{BKM2}.}\label{PEss} 
\end{figure}

\begin{figure}[h!]
\center\includegraphics[width=0.48\textwidth]{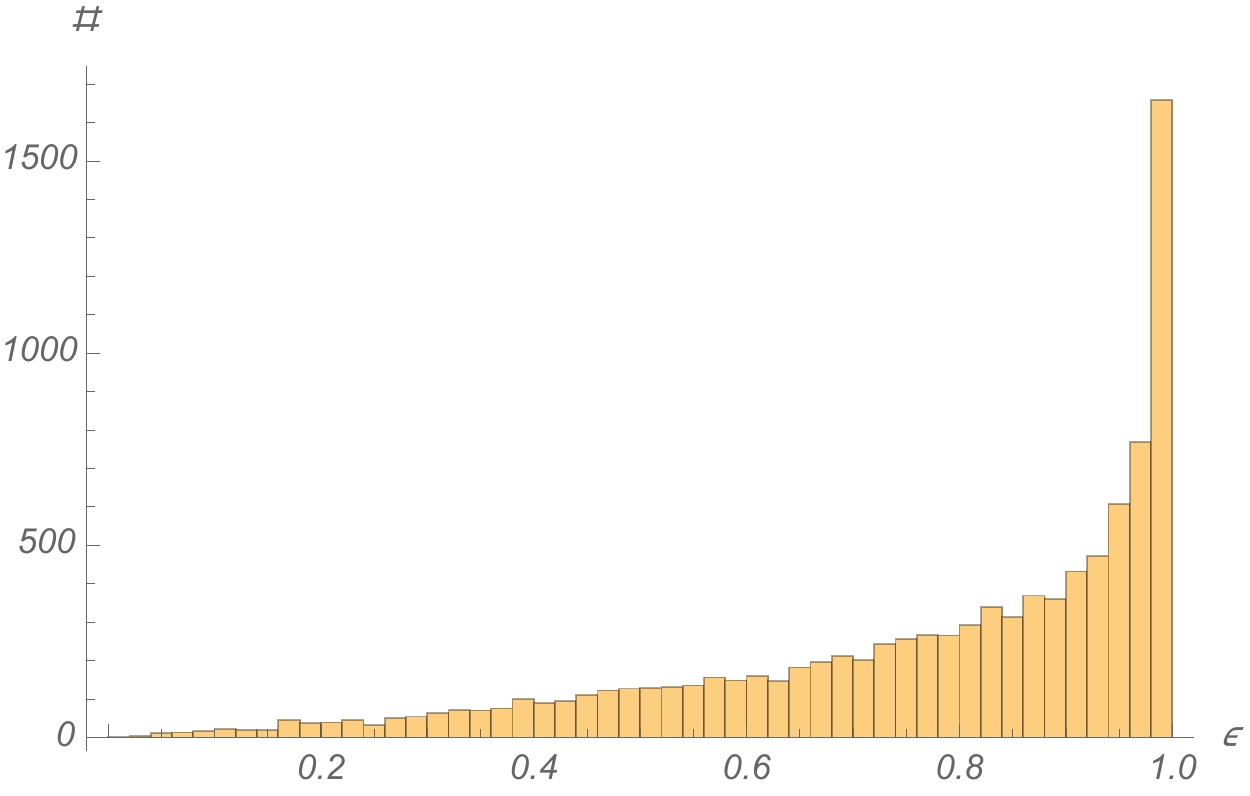}~~~\includegraphics[width=0.48\textwidth]{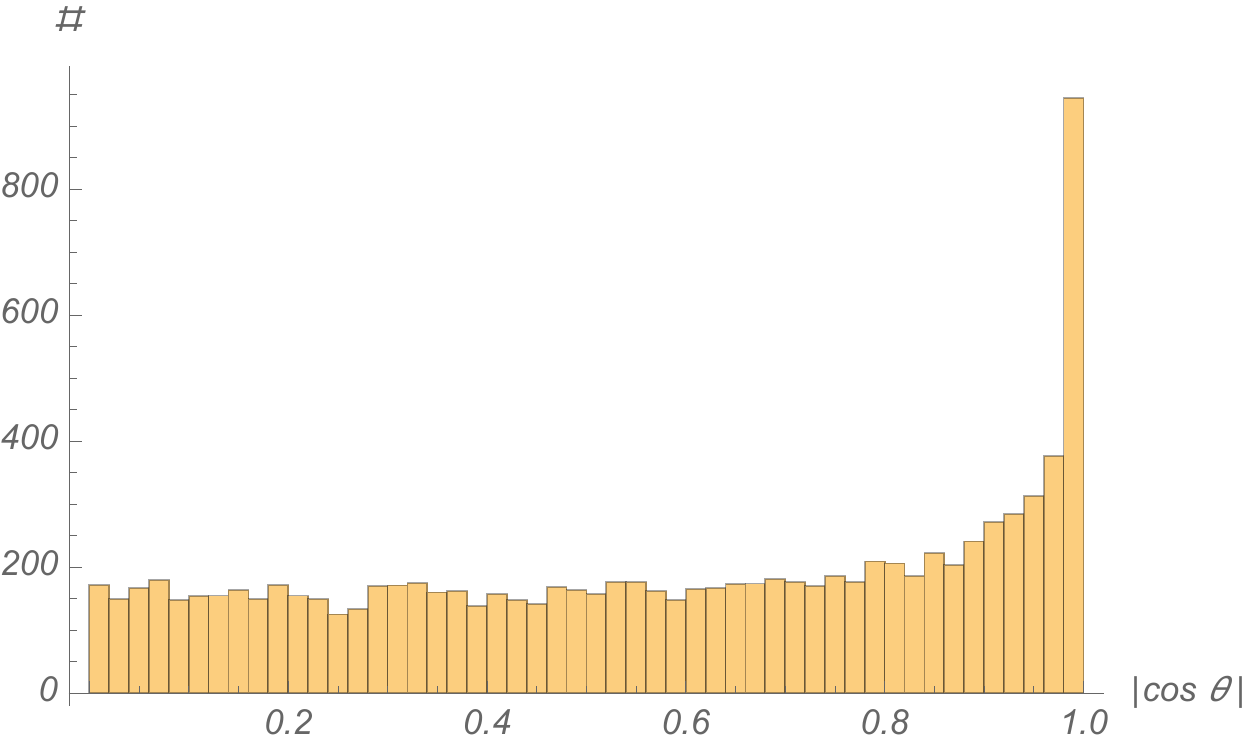}
\caption{\small Distribution of the asymptotic values of $\epsilon$ and $|\cos \theta|$ for numerical simulations of the equal-mass 3-body problem obtained by choosing mid-point data at random on the projectivized cotangent bundle $PT^*{\sf S}$ with uniform distribution according to the natural measure on $PT^*{\sf S}$; see Appendix~A).  Our code performs a simulation starting from the Janus point until sufficiently late times that a Kepler pair is likely to have formed. Then it identifies the pair by proximity (the threshold is that $r_\st{12}^2 < 10^{-4} r_{a3}^2$, $a=1,2$), and it calculates the eccentricity through the formula $\epsilon = \sqrt{1 + \frac{2}{m_\st{red}} E_\st{pair} \| {\bf L}_\st{pair}\|^2}$, where $E_\st{pair}$ and ${\bf L}_\st{pair}$ are the energy and the angular momentum of the pair, and $m_\st{red}$ is the reduced mass of the pair. At the same time, our code calculates the cosine of the angle between the direction of the isolated particle and the semimajor axis of the pair. The latter is determined by looking for local maxima of the pair length $r_{12}$.} \label{Histograms}
\end{figure}

We hold that falsification of our theory of the statistics of mid-point data appears to be possible through the predictions it makes for the asymptotic behaviour of our model. Consider first the 3-body problem (which is planar if ${\bf L}=0$), whose typical solutions we describe in Fig.~\ref{PEss} and caption. Except for zero-measure data at $\jan$, there are always a Kepler pair and an escaping inertial particle in both $t\rightarrow\pm\infty$ limits. In these limits, two distinguishing scale-invariant quantities stabilize: the eccentricity $\epsilon$ of the pair,  and the angle $\theta$ between the semimajor axis of the orbit of the pair and the coordinate vector of the isolated particle relative to the centre of mass of the 3-body system. Numerical calculations for mid-point data taken randomly on the Janus surface give characteristic distributions of the asymptotic values of $\epsilon$ and $\theta$, as shown in Fig.~\ref{Histograms}.

\begin{figure}[h!]
\center\includegraphics[width=0.7\textwidth]{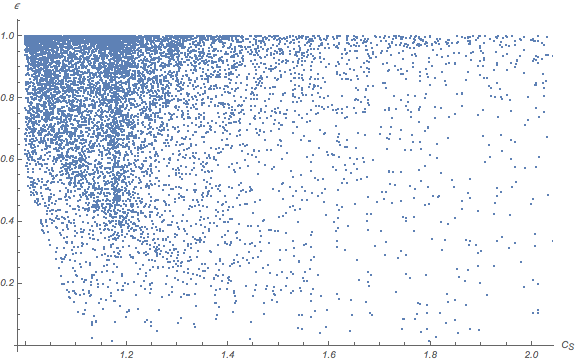}
\caption{\small Correlation between the asymptotic values of $\epsilon$ and the Janus-point values of  $C_\st{S}$. This plot is the result of $\sim 5000$ simulations.} \label{CorrelationEpsilonCs}
\end{figure}

We can go further and consider the correlations of the asymptotic values of $\epsilon$ and $\theta$ with the value of $C_\st{S}$ at the Janus point (which is related to the solution entaxy $\mathcal E_\st{sol}$). In the case of the eccentricity $\epsilon$, we obtain the plot in Fig.~\ref{CorrelationEpsilonCs}.

\begin{figure}[h!]
 \center\includegraphics[width=0.48\textwidth]{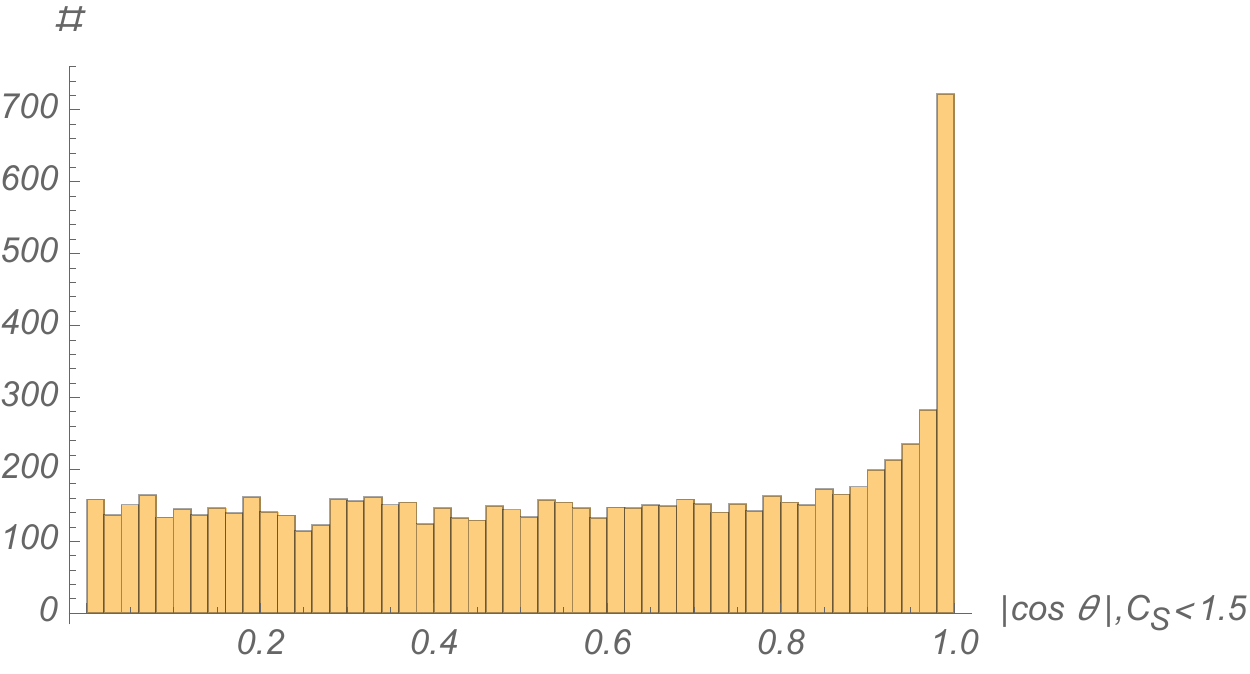}~~~\includegraphics[width=0.48\textwidth]{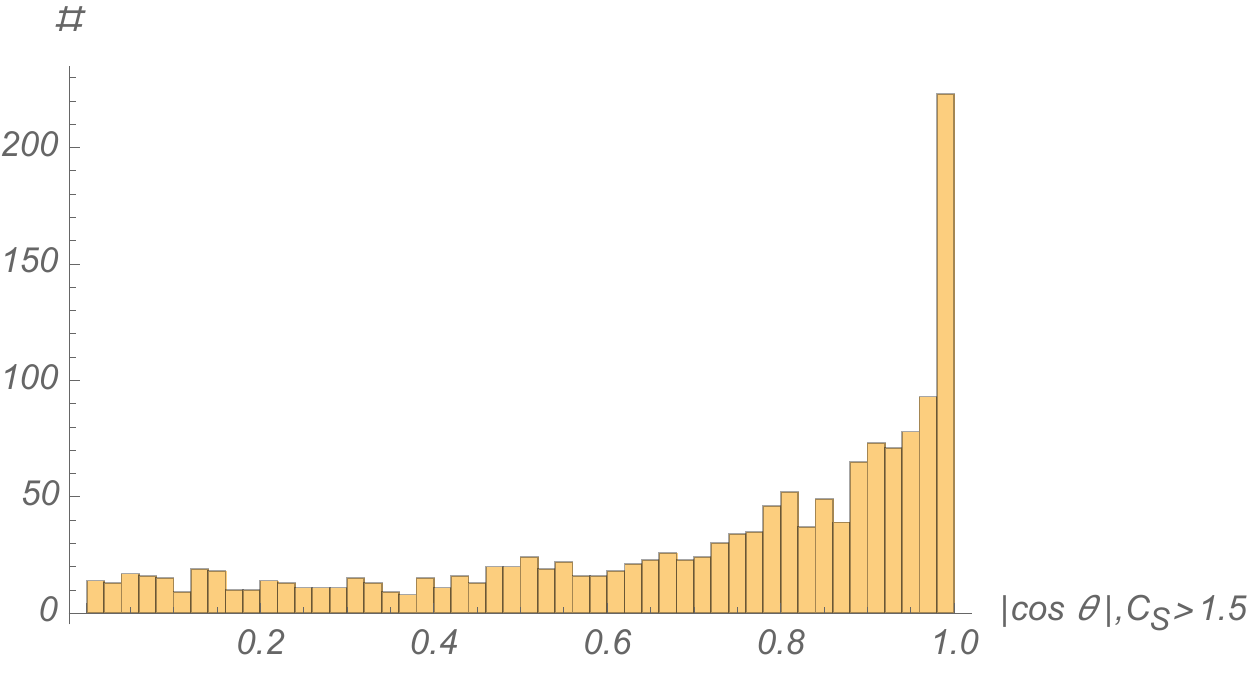}
\caption{\small Distribution of the asymptotic values of $|\cos \theta|$ for solutions with $C_\st{S}$ smaller or larger than $1.5 C_\st{min}$ at the Janus surface $\jan$.} \label{HistogramsSmallerLarger}
\end{figure}

There is clearly some correlation: if we make a histogram of occurrences of values of $\epsilon$ and $C_\st{S}$  `cut out' vertically a some small interval of values of $C_\st{S}$, we obtain a different distribution if the interval is taken around $C_\st{S} \sim 1.1 \, C_\st{min}$ or around  $C_\st{S} \sim 1.6 \, C_\st{min}$. This is more clearly seen in the case of $\theta$: we can plot the histogram of occurrences of $\theta$ at values of $C_\st{S}$ smaller or larger than $1.5 C_\st{min}$, and we get two very different distributions, as shown in Fig.~\ref{HistogramsSmallerLarger}. 

At small Janus-point-complexities, we get a distribution with a peak at $\theta =0$ and a plateau  for any other value of $\theta$. For large complexities, on the other hand, the small values of $\theta$ are penalized in favour of $\theta = 0$. This corresponds to an, as yet admittedly weak, prediction: since small values of $C_\st{S}$ at the Janus point are statistically favoured, our model predicts a distribution of $\theta$ more like the one on the left in Fig.~\ref{HistogramsSmallerLarger}. If, in our hypothetical 3-body universe, values of $\theta$ far from $0$ were measured, that would correspond to evidence against our model.

Of course the predictions that one can make with the 3-body problem are extremely limited and weak, and serve mainly as a proof of principle.

However we conjecture that this dependence of the distribution of asymptotic data on $\mathcal E_\st{sol}$ becomes more marked for some shape variables in the $N$-body problem for large $N$,  for which there will be vastly more scale-invariant quantities that can be correlated with the solution entaxy. If it is correct, then the hypothesis that the solutions with high ${\mathcal E}_\st{sol}$ are the ones most likely to be realized can be tested on the basis of the asymptotic behaviour.\,\footnote{\,In the $N$-body case, the system will in general break up into several clusters and one could, for example, make statistical studies of how the average asymptotic quantities (say 1+3 as opposed to 2+2 decomposition for $N=4$) depend on the typicality of the data at $\jan$. For large $N$, predictions could be made for many observable quantities for typical and atypical mid-point data. Observers in one or the other of the asymptotic limits (as we have noted, they cannot be in both) will always observe an arrow of time and, if our conjecture is correct, be able to establish the extent to which the universe in which they find themselves is typical or atypical. Any $N$-body universe that is maximally typical at $\jan$ will, by virtue of being so, evolve away from $\jan$ to states which are highly atypical and carry information confirming the maximal typicality at $\jan$. This is exactly opposite to the behaviour mandated by the second law of thermodynamics, according to which atypical states evolve to typical ones. Obviously, we must explain the discrepancy, to which we now turn.}

\section{\label{eesl}Entropy and Emergence of the Second Law}

Our discussion of typicality of the shape-dynamical description of a gravitationally dominated universe raises some questions in connection with the second law of thermodynamics.
First, entropy is generally understood as a measure of typicality and the second law of thermodynamics states that entropy never decreases. In Sec.~\ref{SSE} we already made what we think is a convincing case for the inadequacy of thermodynamical entropy as a measure of typicality of the state of the whole universe. Typicality on shape space is different from thermodynamical entropy, which depends on dimensionful quantities and can only belong to subsystems of the universe. This is well illustrated by the example of free expansion of an ideal gas (inertial motion) shown in Sec.~\ref{SSE}: evolution does not correspond to an increase in typicality, because as the gas expands correlations are established between positions and momenta that are not present at the Janus point.\,\footnote{\,Moreover, if adiabatically expanding walls which allow the gas to continually re-equilibrate are not present, there is no useful definition of entropy (already Gibbs' entropy is ill-defined).}

In the rest of the present paper up to this point, we showed how our entaxy captures a suitable notion of typicality of the state of the universe and how its tendency to decrease is linked to the gravitational arrow of time we identified in \cite{BKM3}: it is a consequence of the presence of dynamical attractors in shape space (see Fig.~\ref{PEss}). We believe that this resolves the tension between typicality and the arrow of time \cite{CarrollBook,SW,Davies}, and also opens the way for `retrodiction' of the likelihood of past conditions given a knowledge of  the present ones. This was deemed impossible, for example, by Shiffrin and Wald \cite{SW}, in a universe governed by the second law. The example given in Sec.~\ref{OurTheoryIsPredictive} shows this is in principle possible.

What we have obtained so far is in not in tension with the standard formulations of the second law of thermodynamics. On the contrary, the growth of thermodynamical entropy will be seen to emerge as a characteristic pattern in all solutions of our model. 
In fact, we now want to examine the behaviour of the subsystems as they form progressively and become ever better isolated from the rest of the universe. We will argue that one can assign to them an entropy, defined in the conventional way, which is initially low and then increases. Thus, we have two different kinds of entropy: one for the universe (entaxy) and one for the subsystems that form within it: while the former decreases, the latter increases, matching the standard formulation of the second law of thermodynamics.

\subsection*{Entropy in self-gravitating systems}

The notion of entropy we refer to is the \emph{microcanonical entropy}, which measures our ignorance of the microscopic state of a system. It is mathematically realized by the Boltzmann relation $S = k \log W$, where $W$, in the formulation due to Planck, is the volume of phase space occupied by the states that are compatible with some prescribed values of the state functions (the energy above all). This is, according to Gross \cite{gross2005microcanonical}, still today ``the deepest, most fundamental, and simplest definition of entropy''. Microcanonical entropy is suitable for the description of systems with phase transitions or non-extensive systems with long-range interactions, for which the canonical Boltzmann--Gibbs statistics fails\,\footnote{\,The specific heat of any system described by a canonical ensemble will be positive-definite, $c_\st{V} = \beta^2 \langle (E-\langle E \rangle)^2\rangle$, therefore this ensemble cannot describe systems with negative specific heat. More physically, the canonical ensemble is supposed to describe a system in contact with a thermal bath. If the system has negative specific heat, it will not come into equilibrium with the bath because in such a situation heat flows from hot bodies to cold ones, in violation of Clausius' formulation of the second law.} \cite{padmanabhan1990statistical} and the thermodynamical limit is often singular \cite{gross2005microcanonical}.
In particular, the microcanonical entropy is the only option to describe self-gravitating systems \cite{votyakov2002microcanonical,padmanabhan1990statistical}, because they have a negative specific heat.\,\footnote{\,A negative specific heat is a common feature of all long-range potentials $V \sim r^{-\alpha}$, $\alpha < 2$. This can be proven assuming that the system under consideration is (meta-)stably bound, so that the dilatational momentum is constant on average \cite{pollard1966mathematical}. Then the virial theorem holds, $2 \langle T_\st{k} \rangle \simeq - \alpha \langle V \rangle$, and one can write $\langle E \rangle \simeq  (1-\frac 2 \alpha) \langle T_\st{k} \rangle$, which gives a negative specific heat $c_\st{V} = \frac{\partial \langle E \rangle}{\partial\langle T_\st{k} \rangle}$ if $\alpha<2$.}
 
Microcanonical entropy is not well defined in Newtonian self-gravitating systems unless both a UV and an IR cutoff are imposed to ensure that the hypersurface of constant energy has a finite volume. In fact one needs an IR cutoff, for example a reflecting wall surrounding the system; otherwise the configuration part of the constant-energy surface is non-compact (one can always put a particle arbitrarily far away with a small enough kinetic energy in such a way that the total energy does not change). Also necessary is a UV cutoff preventing Newton's potential from diverging when two particles come arbitrarily close to each other. This can be realized either with a repulsive hard core or with a flattening of the potential below a certain distance. Either way, the purpose of this cutoff is to introduce a minimum for the potential energy, which, in turn, implies a maximum for the kinetic energy on the constant-energy hypersurface. This makes the hypersurface compact. 

Now, it is clear that the nature of our toy model of the universe would be deeply changed by putting it inside a box with reflecting walls. Therefore an IR cutoff cannot be imposed and microcanonical entropy cannot be defined for the whole universe. On the other hand our model spontaneously produces bound subsystems which are effectively `boxed' by their own gravitational potential well for a certain time until a particle escapes, leaving a smaller system that, in its turn, is effectively boxed. For these systems, once they differentiate from the rest of the universe, it makes sense to speak about a microcanonical entropy and an IR cutoff. Then, for these subsystems, all the well-established results on the thermodynamics of self-gravitating systems apply \cite{padmanabhan1990statistical}, and entropy growth is observed as a characteristic tendency of the subsystems to evolve towards local metastable maxima  of the entropy that can escape the UV problem for a considerable time. This fragmentation into subsystems for which the second law holds is an example of the formation of `branch systems', proposed by Reichenbach \cite{Reichenbach,Davies} to solve the puzzle of time-reversal.\,\footnote{\,Despite the utility of Reichenbach's notion, especially as characterized by Davies, we think Reichenbach completely fails to explain the origin of the arrow of time. He assumes that the universe is subject to Poincar\'e recurrences and at no stage even mentions gravity let alone attributes to it a significant role. Davies does not commit himself to the success of Reichenbach's attempt, merely saying ``Reichenbach asserts''.}  
Davies (\cite{Davies}, p.~69) describes branch systems as follows:
\begin{quotation}\small\noindent
Branch systems are regions of the world which separate off from the main environment and exist thereafter as quasi-isolated systems, and usually merge once again with the wider environment after a sufficient time. Examples of this sort are countless, but one will suffice to remove any misunderstanding. When we take an ice cube, and add it to a lukewarm drink, and watch the ice cube melt, the system ice+drink only comes into being \emph{after} the event. It simply did not exist as a quasi-closed system beforehand. Also it will be seen that it really is a quasi- closed system in the sense that the melting process can be perfectly adequately described without recourse to interaction with the outside universe.\normalsize
\end{quotation}
For avoidance of any possible misunderstanding, we note that the human intervention in Davies' example is immaterial; an iceberg falling off Greenland into the Atlantic would serve just 
as well.\,\footnote{\,Albert \cite{albert2009time} is critical of the branch-system concept (p.~88/89), seeing two main problems in it. First, how is it possible ``to decide at \emph{exactly} what moment it was that the glass of water with ice in it first came into being?'' Second, he sees arbitrariness not only in the temporal origin of a branch system but also in its spatial extent. Why not include the room that contains the glass of water with the ice, the building containing the room and so \emph{ad infinitum}?
In fact, a few more words by Davies could have answered these criticisms. First, the moment the ice enters the water is, macroscopically, a well-defined instant that creates locally a manifestly strong departure from equilibrium. Second, one can, with Albert, identify (define) infinitely many branch systems larger in both the temporal and spatial extents that all include the glass, water and ice. These will have boundaries, which need not be physical as long as they are well defined, in both time and space. Now because the number of surface degrees of freedom scale as the square of the linear dimensions while the number within a volume scale as the cube of those dimensions, the positions of the temporal and spatial boundaries have no sensible effect on the melting of the ice. Whatever else may be happening in these extended branch systems, the melting of the ice in Davies' minimal one will contribute an entropy increase in all of the extended ones.}

\subsection*{A dynamical derivation of the second law}

We can now describe our task. We have to explain three facts. First, throughout the universe we see evidence for copious formation of branch systems as described by Davies. Second, the overwhelming majority of them evolve in essentially the same characteristic way that can be called increase of entropy; in each such branch system, this growth defines an arrow of time. Third, all the arrows associated with the formation of such branch systems point in the same direction.
These are the facts we must explain, at least in a reasonably realistic toy model.

We will of course attempt to do this by means of systems with Janus-point solutions. Now there are numerous systems described by equations that are time-reversal symmetric and have Janus points, but three we mentioned earlier -- inertial motion, water waves and systems with repulsive gravity -- do not form branch systems. In contrast, attractive gravity does.\,\footnote{\,Attractive gravity is crucial and its role intriguing. One can barely imagine that the universe is like steam in a cylinder -- it is surely unboxed whatever law governs its behaviour. If we are to obtain a finite measure and hence formulate meaningful statistical statements of the kind made possible by the containers always presupposed in historical studies of thermodynamics and statistical mechanics, we cannot avoid quotienting by scale. This move is supported by the fact that within the universe all observations are ultimately of ratios. However, once branch systems, above all Kepler pairs, form through the action of gravity, dimensional units are defined by them to an ever better accuracy. Moreover, and this completes the circle, the gravitational potential of each negative-energy branch system ensures that it remains effectively confined to a finite region of space -- it creates its own box. Note that if and when a particle does escape from the box, it carries away positive energy. This leads to the formation of a new branch system that is even more tightly bound.}

Our immediate aim is to show that a typical universe is perfectly compatible with the formation of branch systems with arrows of time that, on each side of the universe's Janus point, all point in the same temporal direction. This we can do using the classical $N$-body problem.

We will summarize here in some more detail this process of fragmentation and how it gives rise to subsystems for which the second law holds.
\begin{itemize}
\item
According to corollary 1 of \cite{Marchial:1976fi}, a typical solution $N$-body solution at late times (from our perspective, away from the Janus point in either time direction) will fragment into `subsystems' whose centre of mass drifts away from $\Rcm$ linearly in Newtonian time $t$.\,\footnote{\,See \cite{FlaviosSDtutorial,BKM2} for a definition of Newtonian time purely in intrinsic shape-dynamical terms.} Each subsystem consists of several `clusters' which drift away from each other at most as $t^{2/3}$ \cite{Marchial:1976fi}. 
\item
The fact that the system fragments into clusters has the important consequence that the rest of the universe provides units to describe the state of a cluster. To see this, we can, for example, take an effectively isolated Kepler pair as a reference system. We use its mass to define the reference solar mass and use its semimajor axis as the reference astronomical unit and its orbital period as the reference year. Moreover, we can use its center of mass to define a coordinate origin and define a frame by the orbital plane and the semimajor axis of the Kepler pair. It is thus possible to discuss clusters of the $N$-body system using units of length, time and energy. See \cite{FlaviosSDtutorial} for a more detailed description of the procedure for defining reference frames, a size and a time standard from relational/shape data. It is these definitions that enable us to describe clusters in terms of size, separation, linear momentum, angular momentum and energy. With increasing distance from the Janus point, each cluster will therefore be characterized by physical quantities that become progressively better defined and conserved \cite{BKM2}: energy, linear and angular momentum.\,\footnote{\,These asymptotically conserved quantities are a consequence of the existence of asymptotic Euclidean symmetries for clusters (when a cluster is infinitely far from the others it acquires an effective Lagrangian that is invariant under rotations and space and time translations). Noether's first theorem then applies with ever better accuracy.} These conserved quantities can be used as state functions for a statistical-mechanical description in terms of the microcanonical ensemble \cite{gross2005microcanonical}.
\item
The subsystems virialize, as their moment of inertia grows at most like $\sim t^{4/3}$, and a 
necessary and sufficient condition for the strong form of the virial theorem described in \cite{pollard1966mathematical} is that ${\displaystyle \lim_{t\to \infty}} t^{-2} \, I =0$, which is clearly satisfied.\,\footnote{\,In contrast, the universe has ${\displaystyle \lim_{t\to \infty}} t^{-2} \, I = \textrm{const}>0$, which is why we cannot treat it the same way as its clusters.} Then  the second law will be seen to hold, in the sense that coarse-grained descriptions of the system (given, for example, by the mean-field approximation\,\,\footnote{\,The mean-field approximation only considers the one-particle distribution function $f({\bf p},{\bf q})$ and ignores all correlations between particles ($n$-point functions). It can be shown to arise as the dominant contribution to the microcanonical entropy from a saddle point approximation. The entropy $S_\st{B} = \int \textrm d^3p\,\textrm d^3 q\,f\log f$ associated with this coarse graining is the Boltzmann entropy; it shows a clear tendency to grow in the vast majority of solutions, which, starting from generic initial conditions, will evolve towards configurations which maximize $S_\st{B}$ \cite{padmanabhan1990statistical}. This is the sense in which virialized self-gravitating systems satisfy the second law. The complications due to the long-range nature of gravity relate to the non-existence of entropy-maximizing configurations: there are only local maxima (metastable equilibria) and ultimately no global maximum of $S_\st{B}$. As Sanmartin \cite{sanmartin1995macroscopic} puts it 
``a \emph{runaway in entropy} is a common fate of gravitationally bound systems''.}) will assign an increasing Boltzmann entropy to the vast majority of physical solutions of the system \cite{padmanabhan1990statistical}.
\item 
The statistical mechanics of self-gravitating systems is extremely rich and is closely related to that of systems with first-order phase transitions.
The system admits two distinct phases: a hot diffuse gas which can effectively be
considered an ideal gas with no potential energy, and a collapsed core with very
little kinetic energy. These two phases coexist in actual solutions and the process
of evaporating particles from the core into the diffuse component, which has the effect of collapsing it further, is favoured because it is in the direction of growing entropy \cite{padmanabhan1990statistical}.
This is the basis of the so-called \emph{gravo-thermal catastrophe} \cite{LyndenBell-Wood}. 
The configurations that realize \emph{local} maxima of the entropy for a system with given energy and angular momentum possess a rich and interesting phase diagram \cite{votyakov2002microcanonical}.
\end{itemize}

This shows how the overwhelming majority of branch systems formed in our toy model  evolve in essentially the same characteristic way that can be called increase of entropy.

It remains to explain why the various thermodynamic arrows of time present in subsystems of the universe, the existence of which constitutes the experimental basis of the second law, all point in the same direction as the global gravitational arrow. How does it come about that, in our current epoch, we observe such a plethora of low-entropy branch systems in all of which the entropy is increasing in the same direction? How do all these unidirectional low-entropy branch systems arise?

The only widely accepted answer to this question is the \e{past hypothesis}\,\footnote{\,The coining is due to Albert \cite{albert2009time}.} -- that the universe began in the big bang in a state of extraordinarily low entropy.
Our theory of mid-point data in the self-gravitating $N$-body problem suggests that the problem is an artefact of the transfer of a conceptual framework suitable for the study of steam engines, in which the steam is confined and its self-gravity plays no role, to the self-gravitating universe. 

In our framework, the perceived problem with time-reversal symmetry that leads to the `second-law paradoxes' is eliminated. Because observation is necessarily restricted to part of the universe that is either on one or the other side of its Janus point, the effective law which governs the observed half is not time-reversal symmetric but \emph{asymmetric}. Coupled with the fact that gravity is attractive and tends to clump matter, the formation of branch systems that become ever better decoupled from the rest of the universe is a dynamical necessity. The only role that statistics plays in this process relates to the specific parameters of the branch systems. The fact that they form and with time are described ever better by Hamiltonian equations, so that (negative) energy and centre-of-mass angular momentum are conserved more and more accurately, is inescapable.

It now only remains to show that the all-pervasive dynamical arrow of time thus imposed will ensure that the overwhelming majority of the created branch systems will begin life with a relatively low Boltzmann entropy that will then increase with increasing distance from the Janus point. Since realistic self-gravitating systems cannot have eternally stable maximum-entropy states, we merely note that in our scenario virtually all branch systems consisting of enough particles to make statistical arguments justified will form with some entropy and then evolve to states of higher entropy in which they exist for a relatively long period. Globular clusters provide a good example of what we have in mind: they typically consist of $\sim 10^6$ stars and are mostly well fitted by the Michie--King model~\cite{MillionBodyBook} with a core-halo structure. Many of these were formed very early in the universe and have been slowly evaporating ever since.

There is also the issue of why an initially low entropy virtually never decreases as it would if all the velocities of an entropy-increasing branch system were reversed at the time of its formation. Statistically, both velocity directions are equally possible. However, it is now well understood in conventional statistical mechanics~\cite{Davies} that, in the overwhelming bulk of cases, an initial decrease of entropy will almost immediately be transformed into an increase and, moreover, so rapidly as to be effectively unobservable. The reason for this is that the number of accessible microstates decreases very rapidly with the entropy, so that an entropy-decreasing system very soon `finds its way' for purely statistical reasons into a phase-space region of higher entropy.

As regards the other arrows of time, we have already suggested (in footnote~\ref{retarded}), that retarded potentials will be observed on either side of a Janus point. Further, it is widely accepted that the psychological arrow of time is determined by the thermodynamic arrow, so that should be explained if our overall picture is correct. Another important arrow of time is the one associated with `collapse' of the quantum wave function. On this point, we will merely note that the manner in which the information associated with entanglement `leaks into the environment' as a result of decoherence is strikingly similar to thermodynamic `loss of information', so that too may have its origin in a Janus point.

To summarize, the big picture is that, on either side of its Janus point, the decreasing entaxy of the universe will define an unambiguous arrow of time with which the emergent arrows of the overwhelming majority of the branch systems that form will naturally be aligned. We have, of course, said nothing about non-gravitational forces, which supply much `cleaner' examples of entropy growth than the astrophysical examples we have mentioned. However, we would argue that the formation of \emph{any} branch system that then increases its and its environment's entropy always has its ultimate cause in the clustering effect of extaxy-decreasing gravity. It clearly enabled the sun and earth to form, which in turn led to the creation of coal, the invention of steam engines, and the discovery of the second law.

\section{Conclusions}

In the body of the text, we have said virtually nothing about a key question: is our proof-of-principle $N$-body model a good indicator of what can be expected in the framework of general relativity (GR) or some similar relativistic theory? As we have already written about the architectonic similarities between GR and the \nb in~\cite{BKM1, BKM2, BKM3}, we will here only make some salient points. 

\begin{itemize}

\item The initial-value problem of the relational \nbn, in which a point and a tangent vector in shape space determine a solution, is closely analogous to that in vacuum GR, in which a point and direction in conformal superspace determine a solution~\cite{Barbour:new_cspv,BKM1}. It follows that it will be possible to specify scale-invariant (conformally invariant) mid-point data if suitable Janus points are present in the physically relevant solutions of GR. Then our theory of the typicality of universes should be applicable.

\item In fact, Janus points are present in some classes of GR solutions, for example in big-bang--big-crunch solutions. However, in those cases they are at maximum expansion, whereas  in the \nb they are at minimum expansion. Therefore, more promising is what happens at the big-bang. Since overall scale is a gauge degree of freedom in a truly relational theory, we do not need to worry about the scale singularities that are said to indicate breakdown of classical GR. We merely require regular continuation of the conformal 3-geometry through the big-bang to a universe `on the other side', in which its time will have the opposite direction to ours. This is the most immediately relevant issue to resolve.

\item In GR, the three-dimensional conformally invariant Yamabe constant is a quantity closely analogous to our complexity and could therefore play the role of a primary state function \cite{BKM2}.

\item In its broad features, above all in gravitational clustering and the associated formation of branch systems, the history of our universe since it became transparent at the surface of last scattering does seem to be well reflected in the asymptotic behaviour of the \nbn. This is circumstantial evidence in support of our proposal.\,\footnote{\,Taken together, these points suggest that key properties of the microwave background, such as the temperature equality to $\approx 10^{-5}$ in regions that could never have been in causal contact, could have an \e{acausal} explanation in terms of Laplace's principle applied to the mid-point data of a dynamical system with a very large number of degrees of freedom. If the universe does have a Janus point, Fig.~\ref{HistN} suggests that its mid-point data will correspond to a spatially very uniform state with however microscopic fluctations that can be expected to grow in the initial evolution away from the Janus point.We may also mention that \e{all} Janus-point solutions of our self-gravitating model will have an at least weakly expressed emergent second law. The potential value for cosmology of our theory of typicality is in the predictions it will allow for the `initial conditions'\id mid-point data, of the universe.}

\end{itemize}

Besides these points, we see hope that we are on the right track in the sheer naturalness of an unconfined system like the universe (governed by a time-reversal symmetric law) having a Janus point. Nearly a decade before we realized the significance of the one long known to exist in the \nb (since 1772 in the 3-body problem!), Chen and Carroll~\cite{CarrollChen} conjectured the existence of such a point. Moreover, after we realized that solution-determining data could be
specified at such a Janus point, we found that this idea had been partially anticipated by Carroll with his `middle hypothesis'~\cite{CarrollBook}: if the underlying law of the universe is time-reversal symmetric, one could well expect its solutions to be qualitatively symmetric about a `middle'.
We also discovered that Ashtekar and Sloan (footnote~\ref{bounce}) had already noted how solution-determining data could be specified at a quantum `Janus bounce', though apparently without recognition that this would give a one-past--two-futures scenario.

In view of these facts, we find it especially striking that already Boltzmann (or rather his assistant Dr Schuetz)~\cite{boltzmann1895certain} had, 120 years ago, proposed a one-past--two-futures resolution of the arrow-of-time paradox in the context of a universe subject to Poincar\'e recurrences and hence necessarily confined within a `box' (since Poincar{\'e}'s theorem only holds if the accessible phase space has a bounded measure). Is it not strange, given the long-known `Boltzmann-brain' problems~\cite{albrecht2004can} associated with this idea, that the implications of a Universe Unbound were never, it seems, contemplated?

\section*{Acknowledgements}

We thank Alain Albouy, Boris Barbour, Harvey Brown, Alain Chenciner, Henrique Gomes, Pedro Ferreira, Hamish Forbes, Sean Gryb, Richard Montgomery, David Sloan and Lee Smolin for helpful discussions. This work received financial support from the Foundational Questions Institute (fqxi.org) and the Templeton Foundation.

\section*{Appendix A. The measure in the 3-body problem}\label{A}

In this appendix we will explicitly perform the phase space reduction in the case of three bodies. The extended phase space is $\mathbbm R^6$, with coordinates ${\bf r}_1, {\bf r}_2, {\bf r}_3 \in \mathbbm R^3$ for the particle positions, and ${\bf p}^1, {\bf p}^2, {\bf p}^3 \in \mathbbm R^3$ for the momenta.
It is well known that if the angular momentum ${\bf L} = \sum_{a=1}^3 {\bf r}_a \times {\bf p}^a$ is orthogonal to the plane identified by the three particles, ${\bf L} \times \left( ({\bf r}_1 -{\bf r}_3)\times ({\bf r}_2 -{\bf r}_3)\right) =0 $, then the three particles never leave that plane during the evolution.
In our case the angular momentum ${\bf L}$ is zero and therefore the problem is planar. We can then assume that the position and momenta are two-component vectors  ${\bf r}_1, {\bf r}_2, {\bf r}_3 ,{\bf p}^1, {\bf p}^2, {\bf p}^3 \in \mathbbm R^2$. The degrees of freedom are six, but three are gauge, corresponding to the translations and rotations on the plane of the motion. We have to restrict to the hypersurface ${\bf P}=L_\perp=0$, where $L_\perp$ is the remaining nonzero component of the angular momentum, and then we have to quotient by the transformations generated by these constraints. It turns out that in this case we can take the `royal road' of explicitly identifying a sufficient number of gauge-invariant degrees of freedom (observables), and perform a coordinate transformation in phase space that separates them from the gauge degrees of freedom, making them orthogonal coordinates.

To deal with translations, we define the mass-weighted Jacobi coordinates:
\begin{equation}
\begin{aligned}
\bm \rho_1 &= \textstyle \sqrt{ \frac{m_1 \, m_2 }{m_1 + m_2} }\left( \mathbf  r_2 - \mathbf  r_1 \right), \\
 \bm \rho_2 &=\textstyle \sqrt{ \frac{m_3 \, (m_1+m_2) }{m_1 + m_2 + m_3} } \left({\bf r}_3 - \frac{m_1 \, {\bf r}_1 + m_2 \, {\bf r}_2}{m_1+m_2} \right),\\
 \bm \rho_3 &= \textstyle \frac{1}{\sqrt{m_1+m_2+m_3}} \left( m_1\, \mathbf  r_1 + m_2 \, \mathbf  r_2 + m_3 \, \mathbf  r_3 \right).
\end{aligned}
\end{equation}
The transformation to them is linear and invertible,
\begin{equation}
\bm \rho_a = {M_a}^b \, {\bf r}_b \,,\qquad  \det M = \sqrt{m_1 m_2 m_3},
\end{equation}
and it is therefore easy to find a canonical extension for it:
\begin{equation}
\bm \kappa^a = {(M^{-1})^a}_b \, \mathbf p^b = {((M^{-1})^\st{T})_b}^a \, \mathbf p^b.
\end{equation}
The inverse transformation is
\begin{equation}
{\bf p}^a = {M^a}_b \, \bm \kappa^b = {(M^\st{T})_b}^a \,\bm \kappa^b.
\end{equation}
Note that the inverse matrix is 
\begin{equation}
{(M^{-1})^a}_b =\textstyle \frac{1}{\sqrt{m_1+m_2+m_3}} \left(
\begin{array}{ccc}
 -\sqrt{\frac{m_1 \left(m_1+m_2+m_3\right)}{m_2 \left(m_1+m_2\right)}} &  \sqrt{\frac{m_1 \left(m_1+m_2+m_3\right)}{m_2 \left(m_1+m_2\right)}} & 0 \\
 -\sqrt{\frac{m_3}{m_1+m_2}}  & -\sqrt{\frac{m_3}{m_1+m_2}} & \sqrt{\frac{m_1+m_2}{m_3}} \\
 1 & 1 & 1 \\
\end{array}
\right)\,
\end{equation}
and has a constant column. It is the column of $\bm \kappa^3$,
\begin{equation}
\bm \kappa^3 = \frac{1}{\sqrt{m_1+m_2+m_3}} \sum_{a=1}^3 \mathbf p^a \approx 0,
\end{equation}
which is therefore proportional to the momentum constraint and vanishes. The coordinates
$\bm \rho_3$ are gauge; they are the coordinates of the center of mass.
The other two momenta are
\begin{equation}
\bm \kappa^1 =  \frac{m_1 {\bf p}^2-m_2 {\bf p}^1}{\sqrt{m_1 m_2 \left(m_1+m_2\right)}}\,, \qquad \bm \kappa^2 =\frac{\left(m_1+m_2\right) {\bf p}^3-m_3 ({\bf p}^1+{\bf p}^2)}{\sqrt{\left(m_1+m_2\right) m_3 \left(m_1+m_2+m_3\right)}}.
\end{equation}
\begin{figure}[b!]
\begin{center}\includegraphics[width=0.7\textwidth]{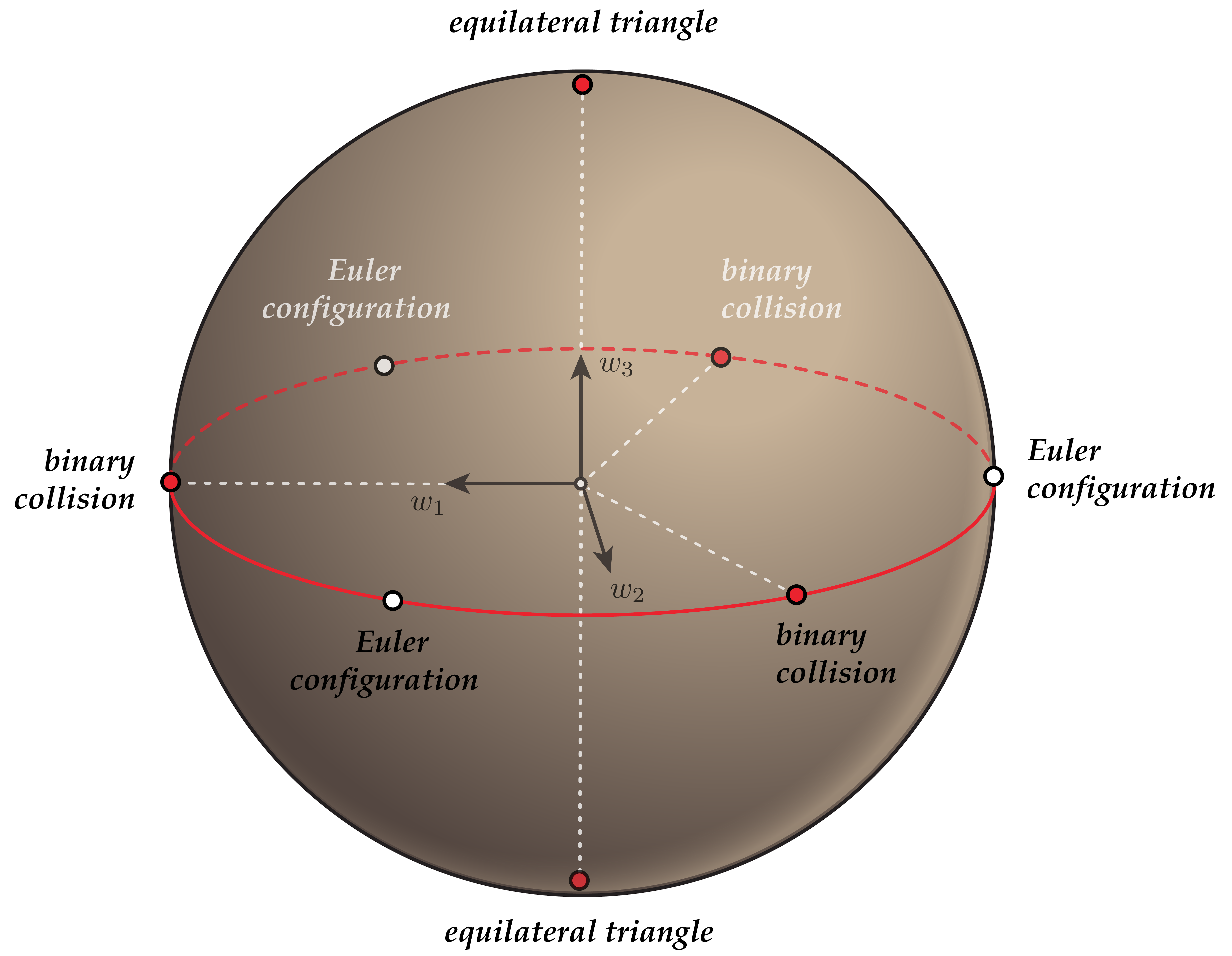}
\end{center}
\caption[The shape sphere]{The shape sphere of the equal-mass 3-body problem. Every point on the sphere
(defined as a constant-$\| \vec w \|$ surface) is a triangle. Points at the same longitude with opposite latitudes correspond to mirror-conjugated triangles. At the poles (the intersections with the $w_3$ axis) we have the equilateral triangles, while on the equator (the red circle, $w_3=0$) we have the collinear configurations. Among them, there are six special ones: three binary collisions (red dots, one of which is on the $w_1$ axis), and Euler configurations (white dots), in which the gravitational force acting on each particle points towards the center of mass and has a magnitude such that, if the system is prepared in rest at one of these configurations, it will fall homothetically (without changing its shape) to a total collision at the centre of mass. The same thing happens at the equilateral triangle (for all values of the masses, as Lagrange showed).}
\label{ShapeSphereFigure}
\end{figure}

The canonical transformation leaves the Poisson brackets invariant:
\begin{equation}
\{ \rho^i_a , \kappa^b_j \} = {\delta^i}_j \,{\delta_b}^a.
\end{equation}
The kinetic term is diagonal in the momenta $\bm \kappa^a$,
\begin{equation}
T =  \sum_{a=1}^3 \frac{\mathbf p^a \cdot \mathbf p^a  }{2 \, m_a} =  \sum_{a=1}^3 \sum_{b,c=1}^3  \frac{{M^a}_b \, {M^a}_c   }{2 \, m_a}  \bm \kappa^b \cdot \bm \kappa^c   = \frac 1 2  \sum_{a=1}^3  \bm \| \bm \kappa^a \|^2,
\end{equation}
as is the moment of inertia,
\begin{equation}
I_\st{cm} = \sum_{a=1}^3 m_a \, \|  {\bf r}_a - {\bf r}_\st{cm} \|^2 = \sum_{a=1}^3 \sum_{b,c=1}^3  {(M^{-1})_a}^b \, {(M^{-1})_a}^c   m_a \,  \bm \rho_b \cdot \bm \rho_c  = \sum_{a=1}^3 \| \bm \rho_a\|^2.
\end{equation}
The inertia tensor also takes a particularly simple form:
\begin{equation}
\mathbbm{I}_\st{cm} =\sum_{a=1}^3 m_a\left( \mathbbm{1} \, {\bf r}_a^\st{cm}\cdot{\bf r}_a^\st{cm} - {\bf r}_a^\st{cm}\otimes{\bf r}_a^\st{cm} \right) 
= \sum_{a=1}^2 \left(   \mathbbm{1} \bm \rho^a \cdot \bm \rho^a - \bm \rho^a \otimes \bm \rho^a \right).
\end{equation}

We are left with four coordinates $\bm \rho_1$, $\bm \rho_2$ and momenta $\bm \kappa^1$, $\bm \kappa^2$, and a single angular momentum constraint (the component perpendicular to the plane of the triangle):  
\begin{equation}
L_\perp = \frac{{\bf L} \cdot \left( ({\bf r}_1 -{\bf r}_3)\times ({\bf r}_2 -{\bf r}_3)\right)}{\left\| ({\bf r}_1 -{\bf r}_3)\times ({\bf r}_2 -{\bf r}_3)\right\|} = \sum_{a=1}^2 \, \bm \rho_a \times \bm \kappa^a \approx 0,
\end{equation}
where with the vector product between two 2-dimensional vector we understand a scalar ${\bf a} \times {\bf b} = a_x b_y - a_y b_x$. The coordinates
\begin{equation}
w_1 = \frac 1 2 \left(||\bm \rho_1||^2 - || \bm \rho_2 ||^2 \right), \qquad
w_2 = \bm \rho_1 \cdot \bm \rho_2\,,  \qquad   w_3 = \bm \rho_1 \times \bm \rho_2
\end{equation}
are invariant under the remaining rotational symmetry and therefore give a complete coordinate
system on the reduced configuration space. Notice that $w_3$ changes sign under a planar reflection (changing the sign of one of the coordinates, say $x$, of both ${\bm \rho}_1$ and ${\bm \rho}_2$) while $w_1$ and  $w_2$ remain invariant, and therefore the map $w_3 \to - w_3$ relates triangles conjugate under mirror transformations. This also has the consequence that the $w_3=0$ plane contains only collinear configurations (whose mirror image is identical to the original, modulo a planar rotation). This has nothing to do with 3D reflections (obtained by changing the sign of \emph{all} components of every Euclidean vector). In fact triangles are invariant under such parity transformations, because their parity conjugate is related to the original by a non-planar rotation.

The Euclidean norm of the 3D vector $\vec w  = (w_1,w_2,w_3)$ is proportional  to (half) the square of the moment of inertia \begin{equation}
||\vec w||^2  = \frac 1 4\left(||\bm \rho_1||^2 + || \bm \rho_2 ||^2 \right)^2 = \frac{I^4_\st{cm}} 4,
\end{equation}
so the angular coordinates in the three-space $(w_1,w_2,w_3)$ coordinatize shape space, which has the topology of a sphere~\cite{Mont}.  We call it the \emph{shape sphere}, and in Fig.~\ref{ShapeSphereFigure} we describe its salient features.

The norms of the original Jacobi coordinate vectors can be written as
\begin{equation}
\| {\bm \rho}_1 \|^2 = \sqrt{w_1^2+w_2^2+w_3^2}+w_1
\,, \qquad \| {\bm \rho}_2 \|^2 = \sqrt{w_1^2+w_2^2+w_3^2}-w_1,
\end{equation}
and therefore the full vectors are specified by
\begin{equation}\label{JacobiCoordinatesVsWcoordinates}
\begin{aligned}
{\bm \rho}_1= \sqrt{ \sqrt{w_1^2+w_2^2+w_3^2}+w_1 } \left( \cos(\theta - \delta/2),\sin(\theta - \delta/2)\right)
,\\
{\bm \rho}_2= \sqrt{ \sqrt{w_1^2+w_2^2+w_3^2}-w_1 } \left( \cos(\theta + \delta/2),\sin(\theta + \delta/2)\right),
\end{aligned}
\end{equation}
where $\delta = \arctan \frac{w_3}{w_2}$ is the angle between ${\bm \rho}_1$ and ${\bm \rho}_2$, and $\theta$ is an overall orientation angle which is not rotation-invariant and therefore is not fixed by the specification of the coordinates $w_1,w_2,w_3$.
%
%
%
%
Now we want to find the momenta conjugate to $\vec w$. To do so, we consider the symplectic potential
\begin{equation}
\Theta =  \sum_{a=1}^3 {\bf p}^a \cdot \d {\bf r}_a =
 \sum_{a=1}^3 {\bm \kappa}^a \cdot \d {\bm \rho}_a.
\end{equation}
If we replace ${\bm \rho}_a$ with their expressions in terms of $\vec w$ from Eq.(\ref{JacobiCoordinatesVsWcoordinates}), we get
\begin{equation}
\Theta = z^1 \, \d w_1 +z^2 \, \d w_2 +z^3 \, \d w_3 + {\bm \kappa}^3 \cdot \d {\bm \rho}_3  + L_\perp \, \d \theta,
\end{equation}
where
\begin{align}
&z^1= \textstyle   \frac{ {\bm \kappa}^1 \cdot {\bm \rho}_1 - {\bm \kappa}^2 \cdot {\bm \rho}_2 } {  \| {\bm \rho}_1\|^2 + \| {\bm \rho}_2\|^2}, \nonumber \\
&z^2 = \textstyle \frac{  {\bm \kappa}^1 \cdot {\bm \rho}_2 + {\bm \kappa}^2 \cdot {\bm \rho}_1  }{  \| {\bm \rho}_1\|^2 + \| {\bm \rho}_2\|^2} - \frac 1 2  \frac{{\bm \rho}_1 \times {\bm \rho}_2}{\|{\bm \rho}_1\| \| {\bm \rho}_2\| } \frac {   \| {\bm \rho}_1\|^2 - \| {\bm \rho}_2\|^2 }{  \| {\bm \rho}_1\|^2 + \| {\bm \rho}_2\|^2} \left( {\bm \kappa}^1 \times {\bm \rho}_1 +   {\bm \kappa}^2 \times {\bm \rho}_2\right),  \\
&z^3 = \textstyle\frac{  \| {\bm \rho}_1\|^2  {\bm \kappa}^1 \times {\bm \rho}_2  -   \| {\bm \rho}_2\|^2 {\bm \kappa}^2\times {\bm \rho}_1 }{2  \| {\bm \rho}_1\|^2 \| {\bm \rho}_2\|^2 } - \frac 1 2  \frac{{\bm \rho}_1 \times {\bm \rho}_2}{\| {\bm \rho}_1\| \| {\bm \rho}_2\| } \frac {   \| {\bm \rho}_1\|^2 - \| {\bm \rho}_2\|^2 }{  \| {\bm \rho}_1\|^2 + \| {\bm \rho}_2\|^2}\left(  {\bm \kappa}^1 \cdot {\bm \rho}_1 - {\bm \kappa}^2 \cdot {\bm \rho}_2 \right), \nonumber
\end{align}
%
%
%
%
so we now have a complete canonical transformation from the coordinates (${\bf r}_1$, ${\bf r}_2$, ${\bf r}_3$; ${\bf p}^1$, ${\bf p}^3$, ${\bf p}^3$) to ($w_1$, $w_2$, $w_3$, ${\bm \rho}_3$, $\theta$; $z^1$, $z^2$, $z^3$, ${\bm \kappa}^3$, $L_\perp$). The Poisson brackets in these coordinates are,  as they should be, canonical:
\begin{equation}
\begin{aligned}
\{ z^a , w_b \} = \delta^a{}_b,  \qquad \{ L_\perp , \theta\} = 1,\qquad \{ z^a , L_\perp \} = 0,
 \qquad \{ z^a, z^b\} =0,
\\
\{ z^a ,\kappa^3_j\} = 0,
 \qquad \{ \theta, \kappa^3_j\} =0, 
 \qquad \{ z^a ,\rho_3^j \} = 0,
 \qquad \{ \theta, \rho_3^j\} =0.
\end{aligned}
\end{equation}

In the new coordinates, the kinetic energy decomposes as
\begin{equation}
T = \frac 1 2 \sum_{a=1}^3 \| {\bm \kappa}_a \|^2  =   \| \vec w  \| \left(  \| \vec z \|^2 + \frac{L_\perp^2}{2(w_2^2 + w_3^2)}\right) +  w_1 \left( \frac{w_3 z^2 - w_2 z^3}{w_2^2 + w_3^2}\right) L_\perp + \frac 1 2 \| {\bm \kappa}^3 \|^2,
\end{equation}
and Newton's potential takes the form
\begin{equation}
V_\st{New} = - \sum_{a<b} \frac{ (m_a \, m_b)^{\frac 3 2}(m_a + m_b)^{-\frac 1 2}}{\sqrt{ \| \vec w  \|  -  w_1  \, \cos \, \phi_{ab} -  w_2  \, \sin \, \phi_{ab}}},
\end{equation}
where the three collision longitudes are \cite{FlaviosSDtutorial}
\begin{equation}
 \phi_{12} = \pi, \qquad
\phi_{13}  = - \phi_{23}  =  \arctan \left(  \textstyle 2 \frac{\sqrt{m_1 \,m_2 \, m_3 (m_1+m_2+m_3) }}{m_2 (m_1+m_2+m_3) - m_1 \, m_3} \right).
\end{equation}
If we call $\phi$ the azimuthal and $\psi$ the polar angle on the shape sphere, then $\frac{w_1}{\| \vec w  \|} = \cos \phi \, \cos \psi$ and  $\frac{w_2}{\| \vec w  \|} = \sin \phi  \, \cos \psi$, and on the constraint surface $L_\perp = {\bm \kappa}^3 =0$ the Hamiltonian constraint takes the form
\begin{equation}
H = \| \vec w  \| \,  \| \vec z \|^2 - \frac {C_\st{S}(\psi,\phi)}{ \sqrt{\| \vec w  \|}}  \,, \qquad C_\st{S}(\psi,\phi) = \sum_{a<b} \frac{ (m_a \, m_b)^{\frac 3 2}(m_a + m_b)^{-\frac 1 2}}{\sqrt{ 1  - \cos \psi  \cos (\phi - \phi_{ab}) }} \,,
\end{equation}
where $C_\st{S}$ is the 3-body complexity function.
Finally a short calculation reveals that the dilatational constraint takes basically the same form in the new coordinates:
\begin{equation}
D = \sum_{a=1}^3 {\bf r}_a \cdot {\bf p}^a  = 2 \, \vec w \cdot \vec z + {\bm \kappa}^3 \cdot {\bm \rho}_3 \,.
\end{equation}

We are now ready to specify our measure on shape phase space. We will gauge-fix translations and rotations simply by setting $\theta=0$ and ${\bm \rho}_3 =0$. Moreover the dilatation constraint $D$ and the Hamiltonian constraint $H$ will gauge-fix each other:
\begin{equation}
\sigma = \int  \left|  \text{F-P} \right| \delta^3({\bm \rho}_3) \delta^3({\bm\kappa}^3) \delta(\theta) \delta(L_\perp)  \delta(H) \delta(D) (\d \Theta)^{\wedge3},
\end{equation}
where the Faddeev--Popov determinant reduces to
\begin{equation}
 \left|  \text{F-P} \right| = \left| \{ D , H \} \right| = \left| 2 \| \vec w  \| \,  \| \vec z \|^2 +  \frac {C_\st{S}} {2\sqrt{\| \vec w  \|}} \right|,
\end{equation}
while the delta-function of the Hamiltonian constraint can be written as
\begin{equation}
\delta(H) = \sqrt{\| \vec w  \|}  \left| 2 \| \vec w  \|\,  \| \vec z \|^2 +\frac {C_\st{S}} {2\sqrt{\| \vec w  \|}}  \right|^{-1} \delta\left(\sqrt{\|\vec w\|}-C_\st{S}^{1/3}\| \vec z\|^{-2/3}\right).
\end{equation}
So the product of the two gives
\begin{equation}
 \left|  \text{F-P} \right| \delta(H) = \sqrt{\| \vec w  \|}  \delta\left( \sqrt{\|\vec w\|} - C_\st{S}^{1/3}\| \vec z\|^{-2/3}  \right),
\end{equation}
and we then have
\begin{equation}
\sigma = \int  \sqrt{\| \vec w  \|}  \delta\left(\sqrt{\|\vec w\|} - C_\st{S}^{1/3}\| \vec z\|^{-2/3} \right)  \delta(2\vec w \cdot \vec z) \, \d^3 w \d^3 z.
\end{equation}
If we now perform a last canonical transformation to polar coordinates in shape phase space,
\begin{equation}
\begin{aligned}
&\vec w = w ( \sin \psi \cos \phi, \sin \psi \sin \phi , \cos \psi) \,,  
\\
&z^1 = \frac 1 w \left( \cos \phi  (w  z^w \sin  \psi - z^\psi  \cos  \psi )- z^\phi  \csc  \psi  \sin \phi \right),
\\
&z^2 =   \frac 1 w \left( \sin \phi  (w  z^w \sin  \psi - z^\psi  \cos  \psi )+ z^\phi  \csc  \psi  \cos \phi \right),
\\
&z^3 =   z^w \cos  \psi - \frac 1 w  z^\psi  \sin  \psi,
\end{aligned}
\end{equation}
then
\begin{equation}\displaystyle
\d^3 w \d^3 z = \d w \d \psi \d \phi \d z^w \d z^\psi \d z^\phi, 
~~~
\vec w \cdot \vec z = w \, z^w,
~~~
\| \vec z \|^2 = (z^w)^2+ \frac{(z^\psi)^2+(z^\phi)^2 \csc^2 \psi}{w^2}
\end{equation}
and
\begin{equation}
\sigma =  \int \delta\left(\sqrt{w} - C_\st{S}^{1/3} ((z^w)^2+  ((z^\psi)^2+(z^\phi)^2 \csc^2 \psi)w^{-2})^{-1/3} \right)  \delta(z^w) \, \d \sqrt{w} \d \psi \d \phi \d z^w \d z^\psi \d z^\phi,
\end{equation}
which finally leads to
\begin{equation}
\sigma =   \d \psi \d \phi \d z^\psi \d z^\phi.
\end{equation}
Notice that we would have got the same result had we used any other gauge-fixing of $D$ than $H$, for example $I_\st{cm} = \textrm{const}$.

Our $\sigma$ is the natural measure on $T^*\shs$, the cotangent bundle to the shape sphere. It is not integrable, because $z^\psi$ and $z^\phi $ do not have a compact domain of definition (or, in other words, the cotangent space to a point in shape space is not compact). The last step is to recall that solutions with the same mid-point data except for a rigid rescaling of all shape momenta $(z^\psi,z^\phi) \to \textrm{const.} (z^\psi,z^\phi)$ are all similar and therefore indistinguishable on shape space. Then there is an equivalence class on $T^*\shs$ of physically indistinguishable mid-point data, and we would like to define our measure on the quotient of $T^*\shs$ with respect to that equivalence relation, which is $PT^*\shs$, the projectivized cotangent bundle to $\shs$.

Now there is no unique way of inducing a measure on $PT^*\shs$, but there is a distinguished one. Consider the scale-invariant mass metric  on the extended configuration space:
\begin{equation}
\d s^2 = \sum_{a=1}^3 m_a \frac{\d {\bf r}_a \cdot \d {\bf r}_a}{I_\st{cm}} = \sum_{a=1}^3 \frac{\d {\bm \rho}_a \cdot \d {\bm \rho}_a}{\| {\bm \rho}_1\|^2 +\| {\bm \rho}_2\|^2 }\,.
\end{equation}
This metric induces a quotient metric on $\shs$ \cite{FlaviosSDtutorial}, 
\begin{equation}
\d s^2 = \d \psi^2 + \sin^2 \psi \, \d \phi^2,
\end{equation}
which, in turn, induces a norm on the associated cotangent bundle $T^*\shs$: 
\begin{equation}
\| \vec z \|^2 = (z^\psi)^2+(z^\phi)^2 \csc^2 \psi.
\end{equation}
We may then use the space of unit-norm covector fields $(z^\psi)^2+(z^\phi)^2 \csc^2 \psi =1$, 
which is the double cover of  $PT^*\shs$, as a redundant description of our solution space. 
Each solution maps to two conjugate points on this space. The conjugacy relation is simply $(z^\psi,z^\phi) \to (-z^\psi,-z^\phi)$.

So we may define our final measure on the space of solutions as
\begin{equation}
\begin{aligned}
\sigma  &=  \int  \delta \left( (z^\psi)^2+(z^\phi)^2 \csc^2 \psi - 1 \right)  \d \psi \d \phi \d z^\psi \d z^\phi  
\\
&= \int \delta \left( Z^2 - 1 \right) \sin \psi \,  \d \psi \d \phi \d Z \d \chi =
 \sin \psi \,  \d \psi \d \phi \d \chi,
\end{aligned}
\end{equation}
where $Z = \sqrt{ (z^\psi)^2+(z^\phi)^2 \csc^2 \psi}$ and $\chi = \arctan \left( \frac{z^\phi}{\sin \psi \, z^\psi}\right)$. This measure is now finite since $\chi \in [0,2\pi)$.

\bibliographystyle{utphys}

\bibliography{bibEntropy3}

\end{document}